\documentclass[12pt,preprint]{aastex}

\newcommand{\hcop}{{HCO$^+$}}
\newcommand{\nh}{{N$_2$H$^+$}}

\slugcomment{To appear in ApJ}

\shorttitle{Large Scale Survey of NGC\,1333}
\shortauthors{Walsh et al.}

\begin{document}


\title{A Large Scale Survey of NGC\,1333}


\author{Andrew J. Walsh and Philip C. Myers}
\affil{Harvard-Smithsonian Center for Astrophysics,
60 Garden Street, Cambridge, MA, 02138, USA}
\author{James Di Francesco}
\affil{Herzberg Institute of Astrophysics, National Research Council 
of Canada, 5071 West Saanich Road, Victoria, BC V9E 2E7, Canada}
\author{Subhanjoy Mohanty, Tyler L. Bourke, Robert Gutermuth and David Wilner}
\affil{Harvard-Smithsonian Center for Astrophysics,
60 Garden Street, Cambridge, MA, 02138, USA}
\email{awalsh@cfa.harvard.edu}



\begin{abstract}
We observed the clustered star forming complex NGC\,1333 with the BIMA
and FCRAO telescopes in the transitions HCO$^+$(1--0) and N$_2$H$^+$(1--0) over
an $11'\times11'$ area with resolution $\sim10''$ (0.015\,pc).
The \nh~emission follows very closely the submillimeter dust continuum
emission, while HCO$^+$ emission appears more spatially extended and also
traces outflows. We have identified 93 \nh~cores using the {\sc CLUMPFIND}
algorithm, and we derive \nh~core masses between 0.05 and 2.5\,M$_{\odot}$,
with uncertainties of a factor of a few, dominated by the adopted \nh~abundance.
From a comparison with virial masses, we argue that most
of these \nh~cores are likely to be bound, even at the lowest masses,
suggesting that the cores do not trace transient structures, and 
implies the entire mass distribution consists of objects that can potentially form stars.
We find that the mass distribution of \nh~cores resembles the field star
IMF, which suggests that the IMF is locked in at the pre-stellar stage of
evolution. We find that the \nh~cores associated with stars identified
from Spitzer infrared images have a flat mass distribution. This might be
because lower mass cores lose a larger fraction of their mass when forming
a star.  Even in this clustered environment, we find no evidence for
ballistic motions of the cores relative to their lower density surroundings
traced by isotopic CO emission, though this conclusion must remain tentative
until the surroundings are observed at the same high resolution as the \nh.

%
\end{abstract}




\section{Introduction}

Stars form predominantly within clusters inside turbulent molecular cores 
\citep{clarke00}, but there is no generally accepted theory 
of how this process occurs \citep{elmegreen00}. This deficiency is due 
in part to the lack of observational characterization of processes involved 
in cluster formation, given the wide range of scales that must be sampled 
simultaneously to provide a meaningful picture (eg., 0.01--1.0\,pc).
Much recent work has focused on observations of isolated star formation.
This is easy to understand as a star forming in isolation is not affected
by any neighbors. However, in a clustered region, we usually see stars at
different stages of formation, all within close proximity to each other,
implying that there may be an effect of the older stars on the evolution
of the youngest cluster members. An excellent example of this is NGC\,1333,
where Class 0, I, II and III protostars exist, almost certainly along with
prestellar cores. Trying to understand such a complicated region is much more
difficult, but is a worthwhile pursuit because most stars form in clustered
environments where these confusing factors appear. Thus, in order to
understand the typical mode of star formation, it is necessary to understand
star formation in a cluster such as NGC\,1333.

\begin{figure}
\plotone{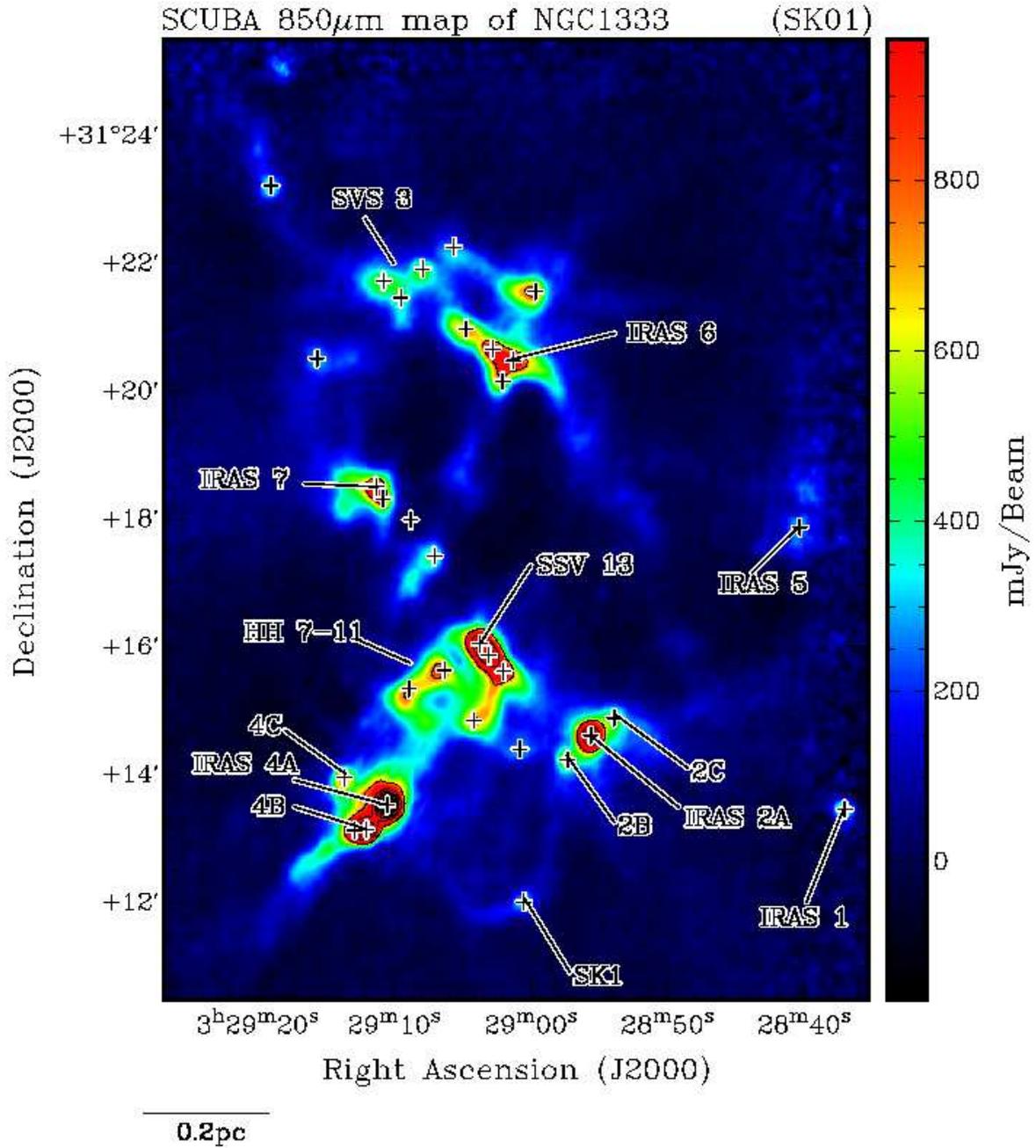}
\caption{SCUBA 850$\mu$m dust continuum emission (SK01) in NGC\,1333.
Contours are 10, 20, 30, ... 90\% of the peak 6.68\,Jy/beam.
Plus symbols indicate the positions of SCUBA continuum sources
identified by SK01.}
\label{scuba}
\end{figure}

NGC\,1333 is a reflection nebula in Perseus which is associated with a molecular cloud,
at a distance of 300\,pc \citep{dezeeuw99}.
It is also associated with two young star clusters identified in the near-infrared
\citep{aspin94,lada96}. Furthermore, it appears that a significant proportion of
the stars seen in the near-infrared are pre-main sequence \citep{aspin94,aspin03}.
Figure \ref{scuba} shows the 850\,$\mu$m dust continuum emission towards NGC\,1333,
mapped by Sandell \& Knee (2001; hereafter SK01).
Plus symbols mark the positions of dust continuum peaks, established by
SK01.
The majority of the emission is found in the southern
half of NGC\,1333, which includes the IRAS 4 complex, the IRAS 2 complex and the
SSV\footnote{We adopt the SSV designation for this source following \citet{herbig83}}
13 complex. Each complex has previously been identified as a site of multiple star
formation.
NGC\,1333 contains many well known outflows. Probably the best known is the outflow
powered by SSV\,13, or possibly the radio continuum source VLA\,3 \citep{rodriguez97}.
It is associated with the Herbig-Haro objects HH\,7-11 to the SE and extends into the central
cavity of NGC\,1333 to the NW. In addition to this,
there are outflows from both the IRAS\,4 and 2 complexes, as well as IRAS 7 and 8 to the north.

SK01 note that the dust cores in NGC\,1333 appear to have a mass function which is consistent with
the stellar IMF. This has also been found for other regions of star formation such as in Ophiuchus
\citep{motte98,johnstone00} and IRAS 19410+2336 \citep{beuther04}. Recent theories for star
formation (eg. \citet{bonnell01a,bate05}) indicate the stellar initial mass function is created at an
advanced stage of evolution -- during the accretion phase as matter is deposited onto the pre-main
sequence star. Thus, the findings of stellar-like mass functions in prestellar cores is unexpected.
One possible explanation is that the dust cores are transient phenomena and therefore are unlikely
to be the sites of future generations of star formation. In order to decide whether such cores are
transient or self-gravitating, it is necessary to observe cores in molecular line tracers so that
kinematic information may tell us about their internal motions.

\section{Observations and Data Reduction}

\subsection{BIMA Observations}
Observations of NGC 1333 were made with the 10-element Berkeley Illinois Maryland Association (BIMA)
interferometer, at Hat Creek, CA, USA, operating simultaneously at 89 GHz and 93 GHz.
The maps were made by mosaicing together 126 pointings spaced by
$\sim$1$^\prime$.
With a $\sim$2\arcmin~FWHM primary beam at 3\,mm, the resulting close-packed
hexagonal pointing pattern provided full Nyquist sampling across the central
11\arcmin $\times$ 11\arcmin, yielding constant sensitivity across this field.
The field was divided into 2 halves, with the southern half observed mostly
in Fall 2001, and the northern half observed in Spring 2002.  A total
of 16 useable data tracks of 3-11 hours length were obtained over this
period, in both the C- and D-array configurations. Both HCO$^+$ (1--0)
and N$_2$H$^+$ (1--0) were observed in correlator windows of 256 channels each 
0.16\,km\,s$^{-1}$ wide. All 7 hyperfine components of N$_2$H$^+$ (1--0) were observed
in a single window. The remaining correlator windows were combined to
provide 800\,MHz bandwidth for simultaneous continuum observations.

We assume the line rest frequencies of 89.188253\,GHz for
\hcop (1--0) \citep{ulich76} and 93.176258\,GHz for the `isolated'
($F_1F = 01-12$) \nh (1--0) hyperfine component \citep{lee01}.

\subsection{FCRAO Observations}
Observations of NGC\,1333 were made with the
Five College Radio Astronomy Observatory (FCRAO) 14-m telescope on 2002
November 16. The data were taken in an On-The-Fly (OTF)
mode\footnote{http://donald.astro.umass.edu/\~{}fcrao/library/manuals/otfmanual.html}
using SEQUOIA, which is a MMIC multibeam receiver with a 32 pixel array,
arranged as two 4$\times$4 arrays with orthogonal polarisations.
Simultaneous observations were performed in \nh (1--0) and \hcop (1--0).
The entire 11\arcmin $\times$ 11\arcmin~BIMA field was mapped with
full Nyquist sampling. The FCRAO beam FWHM is about 50\arcsec~for these observations.
Typical rms noise values are 0.048\,K for \hcop~and
0.053K for \nh. Average system temperatures for \hcop~were 164\,K and for
\nh~were 174\,K. The spectrometer is a digital autocorrelator with 1024
channels. Our spectral resolution was 25\,kHz, corresponding to about
0.08\,km\,s$^{-1}$. We assumed a conversion factor of 43.7\,Jy/K.

\subsection{Combination of Interferometer and Single Dish Data}
The BIMA and FCRAO data were combined using the BIMA version of {\sc MIRIAD}
(Zhang, Q. {\em private communication}).
The FCRAO data were initially demosaiced into single pointings
corresponding to the BIMA pointings. Each pointing was then deconvolved with the FCRAO
beam. UV points were randomly generated across the UV plane, with UV distances
from 0 to uvmax, where uvmax corresponds to the size of the FCRAO dish
(radius 7\,m). The density of points in the UV plane was chosen such that they
match the density of UV points from the BIMA data. The demosaiced, deconvolved
FCRAO data were then transformed into the UV plane. Only data corresponding
to the randomly generated UV points were recorded. The FCRAO and BIMA UV data were
then combined, imaged and cleaned using standard {\sc MIRIAD} reduction
techniques.

The 1-$\sigma$ rms sensitivities of the cleaned data
are 3 mJy beam$^{-1}$ for the continuum data, 330 mJy beam$^{-1}$ for both
HCO$^+$ and N$_2$H$^+$.  The final synthesized beam FWHMs were $\sim$10\arcsec
($\sim$0.011\,pc) for all data.

\section{Results}

\begin{figure}
\plotone{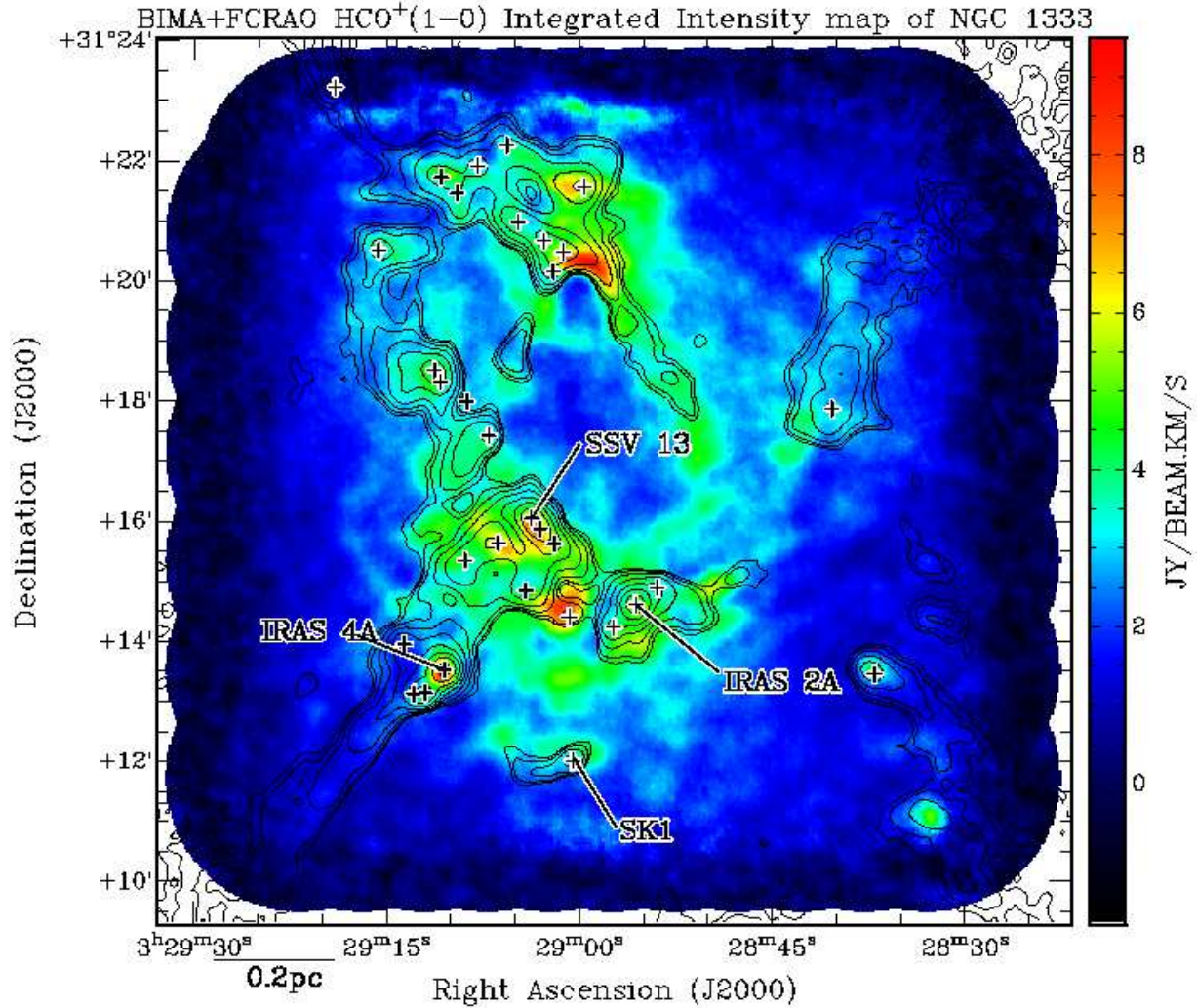}
\caption{HCO$^+$(1--0) integrated intensity map of NGC\,1333. 0.2\,Jy/beam km/s = 1$\sigma$.
Contours are SCUBA 850$\mu$m emission (SK01).
Contour levels are 0.02, 0.04, 0.08, 0.16, ... 5.12\,Jy/beam, where 0.02\,Jy/beam = 1$\sigma$.
Plus symbols indicate the positions of SCUBA continuum sources identified by SK01. HCO$^+$
emission is seen to cover a larger area than submillimeter continuum, and loosely follows
the submillimeter contours.}
\label{hcop}
\end{figure}

Figure \ref{hcop} shows the HCO$^+$(1--0) integrated intensity emission,
compared to the 850$\mu$m continuum emission. Some general correspondence is seen
between the two. For example, many of the peaks seen in the continuum map correspond
to \hcop~peaks, including IRAS\,4A, 4B, 2A, SSV\,13. Also, some large scale structures
show similar morphologies. On the other hand, some significant differences between
the two are noticed: generally the \hcop~emission appears to cover a larger area than the
continuum emission. Although the continuum map has been
filtered spatially due to chopping during observations,
this is probably because the \hcop~transition is more sensitive to
lower column densities of material than the continuum observations. There are also
differences in the distributions due to the effects of outflows which are discussed in
\S\ref{outflows}.

\begin{figure}
\plotone{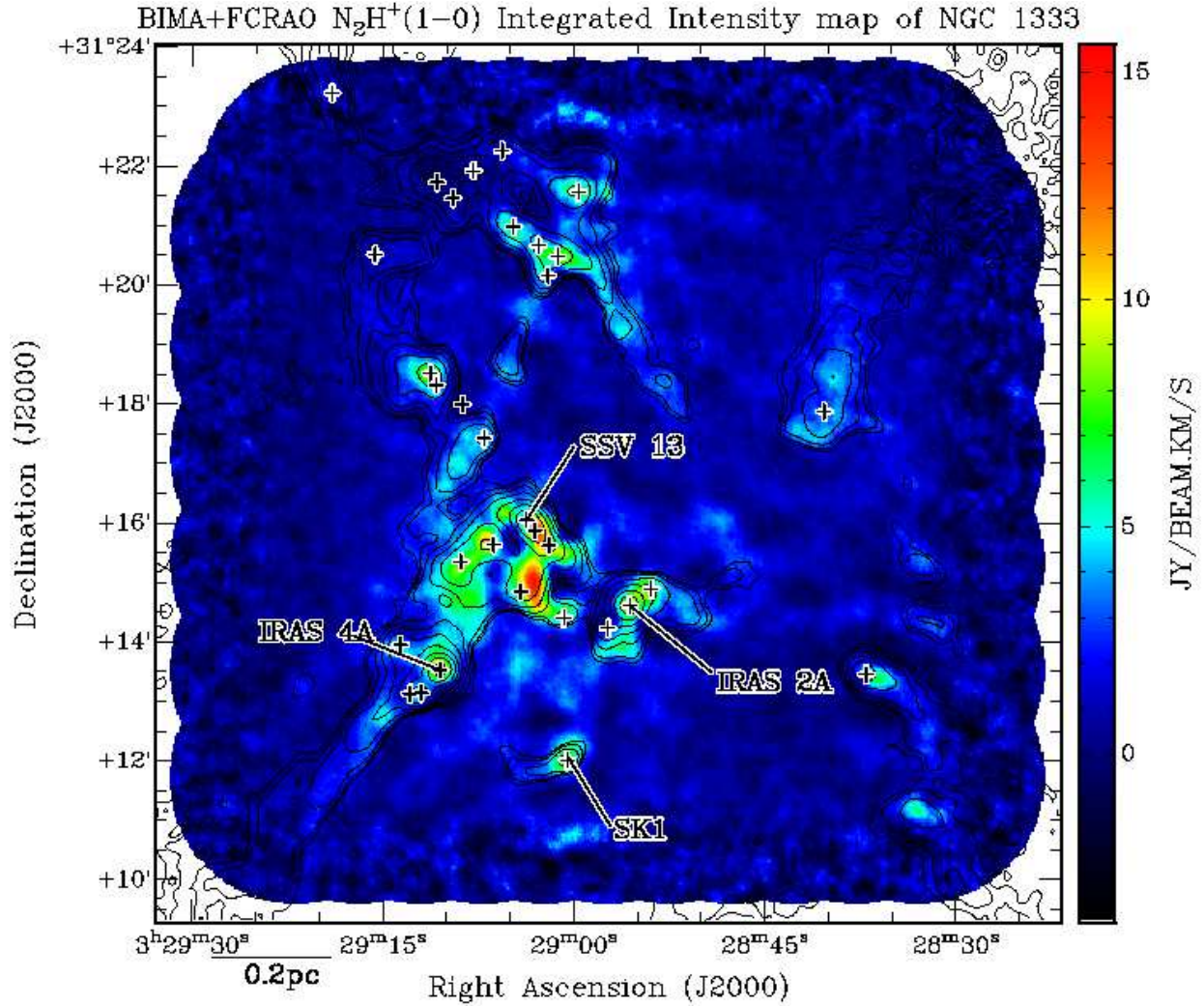}
\caption{N$_2$H$^+$(1--0) integrated intensity map of NGC\,1333. 0.2\,Jy/beam km/s = 1$\sigma$.
Contours are SCUBA 850$\mu$m emission (SK01).
Contour levels are 0.02, 0.04, 0.08, 0.16, ... 5.12\,Jy/beam, where 0.02\,Jy/beam = 1$\sigma$.
Plus symbols indicate the positions of SCUBA continuum sources identified by SK01. N$_2$H$^+$
and submillimeter continuum closely follow each other in both extent and morphology.
Small differences are seen between the two, however (eg. IRAS\,2A/B/C).}
\label{n2hp}
\end{figure}

Figure \ref{n2hp} shows the \nh(1--0) integrated intensity emission, compared to the
dust continuum emission. The distribution of \nh~very closely follows that of the
cold dust emission, much closer than that of the \hcop~emission. This similarity shows
that \nh~is an excellent tracer of cold dust emission, as is known from studies of
isolated cores \citep{tafalla02}. However, some small
differences can be seen in Figure \ref{n2hp}: the brightest spot in the dust continuum
map is IRAS\,4A, which is over twice as bright as anything else in the field of view.
However, the brightest spot in the \nh~map is located an arcminute to the south of
SSV\,13. The IRAS\,2 complex includes three dust continuum peaks 2B, 2A and 2C
running from SE to NW, respectively. Whilst IRAS\,2A is clearly the brightest in both
continuum and \nh~integrated intensity emission, IRAS\,2C is a very weak continuum source
but is almost as bright in \nh~emission as IRAS\,2A. Furthermore, IRAS\,2B has stronger
emission in continuum than IRAS\,2C, but appears almost devoid of any \nh~emission.
\citet{jorgensen04a} have investigated the chemistry of \nh~in this region and conclude that
the apparent differences are due to varying degrees of thermal processing: 2B is the warmest and
oldest source, having destroyed \nh~around it, IRAS\,2C is the youngest and coolest source
with physical conditions that favor the production of \nh, and IRAS\,2A has a temperature and age
somewhere between the two. One other major difference between the continuum and \nh~morphologies
is that near SVS\,3 (to the NE of the field of view), there is
a large region which clearly shows continuum
emission consisting of at least five sources identified by SK01, spanning an area
approximately 0.3\arcmin$\times$0.2\arcmin, but shows no evidence of \nh~emission.

\begin{figure}
\plotone{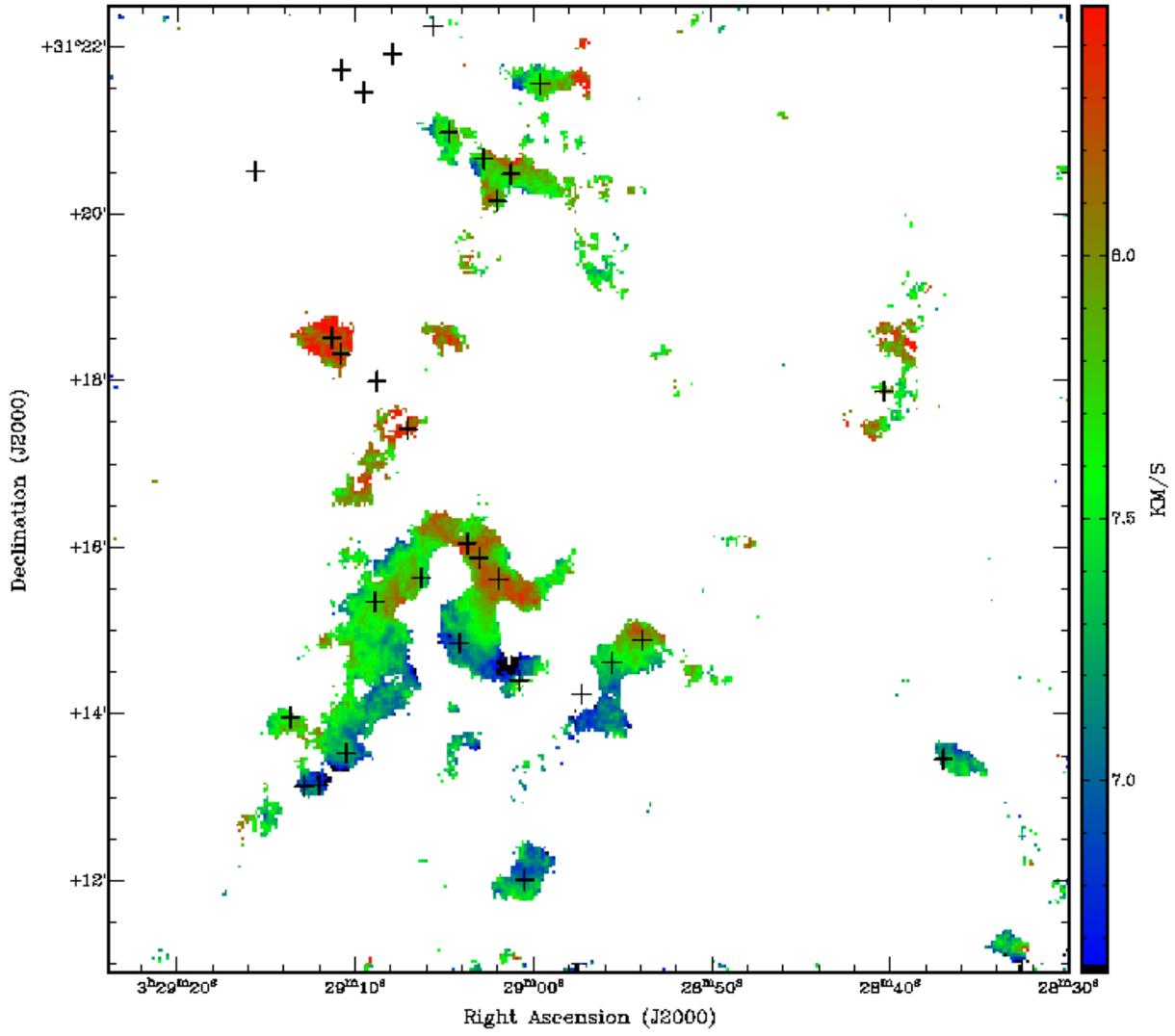}
\caption{\nh~first moment map of NGC\,1333, showing the line of sight motions of the gas.
Plus symbols indicate the positions of SCUBA continuum sources identified by
SK01.}
\label{frames}
\end{figure}

Figure \ref{frames} shows the first moment map of line of sight velocities for \nh . Most of the emission
is found between velocities of 6.5 and 8.5\,km\,s$^{-1}$. Overall, the emission in the southern
part of NGC\,1333 (south of 31$^\circ$ 14$^\prime$) appears blueshifted with respect to emission
in other parts of the region. \nh~emission close to IRAS\,7 appears to be redshifted from the bulk
of the emission. Previously, \citet{ho80} suggested that bulk motions traced by ammonia were indicative of rotation.
However, it appears their data were strongly biased by the redshifted emission associated with
IRAS\,7. If NGC\,1333 were undergoing rotation, then we would expect to see redshifted emission over
a much larger area, rather than being confined around IRAS\,7. We still see a large area to the south
with blueshifted emission, but apart from IRAS\,7, the rest of the gas appears to be moving at approximately
the same velocity. Therefore, the gas traced by \nh~does not show significant signs of large-scale rotation,
although there are clear differences between motions on smaller scales.

%
%
%
%
\subsection{HCO$^+$ outflows}
\label{outflows}
\begin{figure}
\plotone{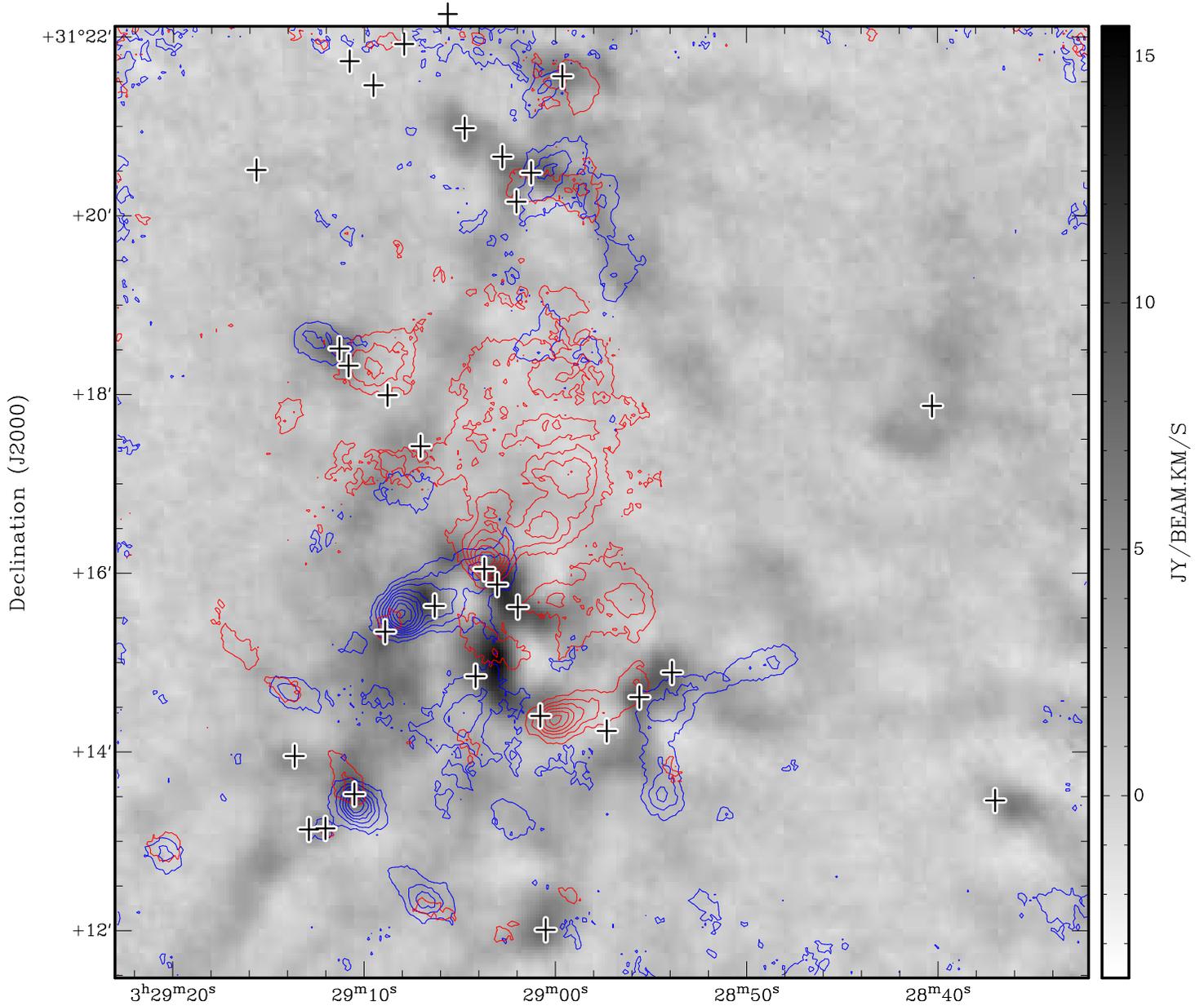}
\caption{\hcop~outflows in NGC\,1333. The greyscale is \nh~integrated intensity. Blue contours represent
\hcop~integrated emission from -5.5 to +5.3\,${\rm km}\,{\rm s}^{-1}$. Red contours represent
\hcop~integrated emission from 10.4 to 16.4\,${\rm km}\,{\rm s}^{-1}$. The lowest contour (both red and blue)
is at 9\,Jy/beam and increase in steps of 4\,Jy/beam to a maximum of 33\,Jy/beam for the red and 41\,Jy/beam
for the blue.}
\label{figoutflows}
\end{figure}

\citet{knee00} have previously described these outflows in detail, however we include Figure
\ref{figoutflows} which shows the blue and red lobes of outflows in the \hcop~map which
shows the outflows at higher resolution than the maps of \citet{knee00}. The blue-shifted lobe
from SSV\,13 (or VLA\,3) is clearly the brightest emission in the field of view.

Apart from the previously detected outflows, we note that the large, redshifted lobe of
emission in the center of the field of view coincides with the central cavity \citep{knee00}.
The central cavity is filled with high velocity gas from outflows and \citet{knee00} suggest
it was created by these outflows. Furthermore,
the red lobe coincident with the central cavity
appears to be bounded by the filamentary structure of dense gas traced by \nh.
We therefore conclude that the outflow(s) responsible for this redshifted lobe fill the
cavity with high velocity gas. It is clear that the bulk of this lobe comes from the outflow
associated with either SSV\,13 or VLA\,3, but the north--south outflow from IRAS\,2A
also projects into this area, as well as the outflow from IRAS\,7. It is even possible that
there is a red-shifted outflow lobe from SSV\,13B, located to the west of SSV\,13 that also
projects into the cavity. These overlapping outflows make interpretation of the red-shifted
emission in this region difficult.

\section{Discussion}
\subsection{Analysis of \nh~cores}
As previously mentioned, the \nh~map very closely follows that of the dust continuum emission.
Therefore we can analyse the density structure of NGC\,1333 by observing the emission in this transition.
Because \nh~is a spectral line, we can also use the velocity information to help discriminate
cores along the same line of sight. The \nh~(1--0) transition has seven hyperfine components
which could potentially be used for better velocity determination of the cores, however,
because most of these hyperfine components are blended, we choose to restrict our analysis
to the isolated hyperfine component. This has the unfortunate effect of reducing our overall
signal from this transition.

{\sc Clumpfind} \citep{williams94} is an algorithm for identifying peaks of emission in a data cube.
It was originally used to identify clumps in single dish spectra.
It works in a way similar to contouring a map: where it finds closed contours a core is
identified. The algorithm also extends the size of the core by including adjoining pixels
down to a detection threshold. If a core is extended until it runs into another core (with
its own separate peak), the algorithm stops extending the core on that front. The outcome
is an interconnected cube of identified cores that appear similar to what would be expected
if one identified cores and core boundaries by eye. Thus, all the real emission above the
detection threshold is categorised as material belonging to some core.

\begin{figure}
\plotone{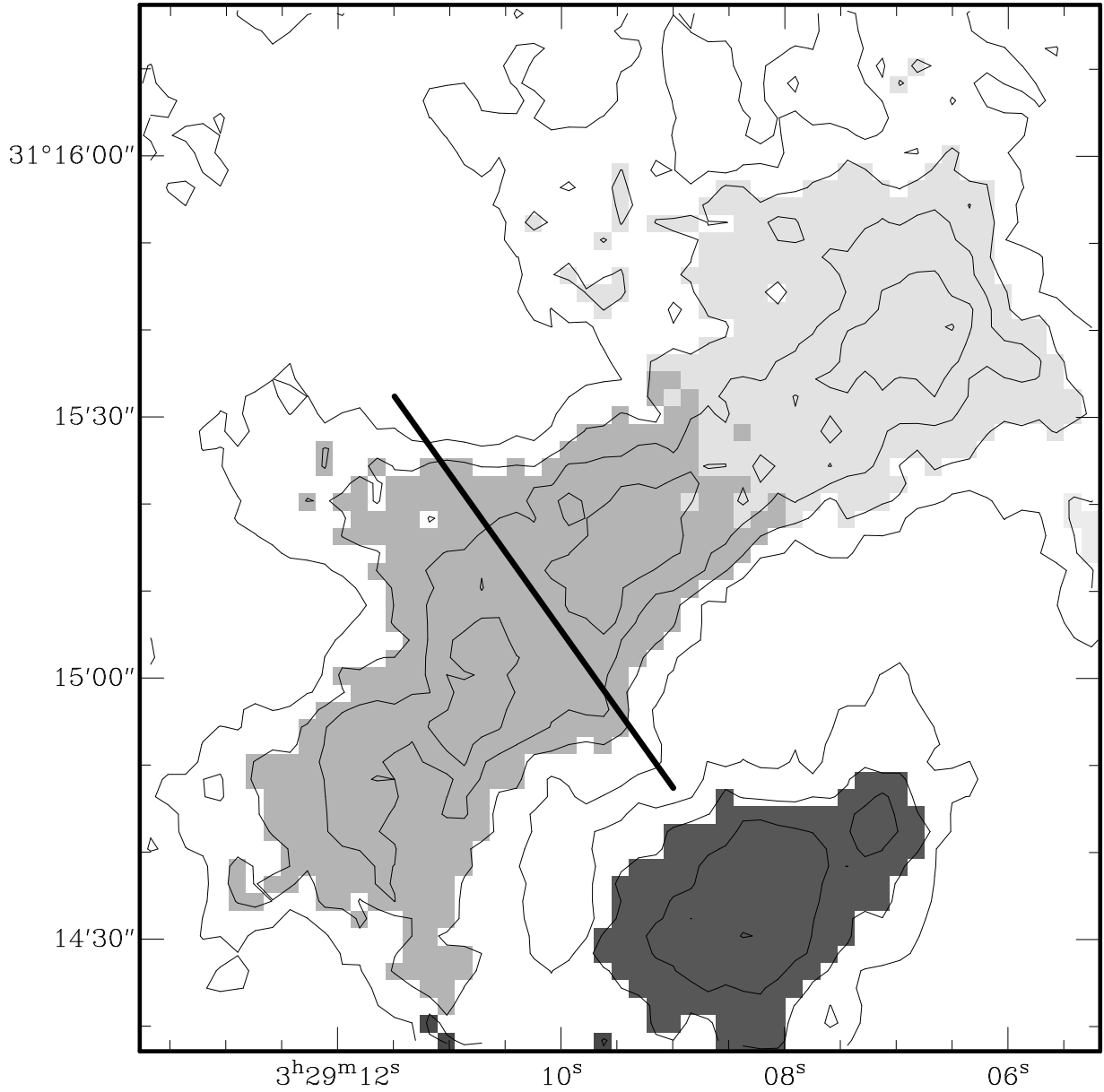}
\caption{{\sc Clumpfind} results for a small portion of the \nh~data cube.
The different shadings represent the identified cores. The contours
are \nh~emission at the same levels as those used by {\sc clumpfind}: 0.244,
0.488, 0.732... 2.196\,Jy/beam. The diagonal line
represents an artificial break in the largest core that should be applied
to separate this core into two individual cores.}
\label{clfind}
\end{figure}

We chose to identify cores using the default setting of searching down to twice the rms noise
level, with spacing between contours also twice the rms noise (0.244\,Jy/beam). We require that
each core must contain at least enough pixels to fill one beam.
We found that sometimes {\sc clumpfind} was inadequate for separating out multiple cores. For example,
Figure \ref{clfind} shows a small portion of one velocity channel of the \nh~data cube,
with identified cores shaded differently. The contours are the \nh~emission and coincide
with the contours used by {\sc clumpfind} to identify cores.
Three cores are identified by {\sc clumpfind} in this region.
However, {\sc clumpfind} failed to break up the largest core along the line shown. We believe that
cores like this should be divided because there are closed contours on both sides of the dividing
line. In this case, we divide the core into two by running {\sc clumpfind} again with a smaller
separation of contours. In this manner, we have avoided limitations of the {\sc clumpfind} program
and we believe we have separated cores in an unbiased manner.
Note that whilst the smallest identified clump, to the south with the darkest shading,
appears to show two closed contours, the smaller peak contains less than the required number
of pixels to fill a beam. Therefore, it is not assigned to be a separate core.

Many of the weakest cores identified by {\sc clumpfind} were found to be noise spikes. We therefore
employed a reality check on all clumps. Because we have data on all seven hyperfine components
of the \nh~line, we checked the \nh~spectrum of the core to make sure all seven components
were visible. This is a very robust test because we identified our cores on a weak hyperfine
component and so it was easy to discern a noise spike from a true \nh~core. Furthermore, we
checked that the morphology of the core identified in the isolated hyperfine component was
similar to that seen in the brighter hyperfine components. As a result we have identified 93 cores.

\begin{figure}
\plottwo{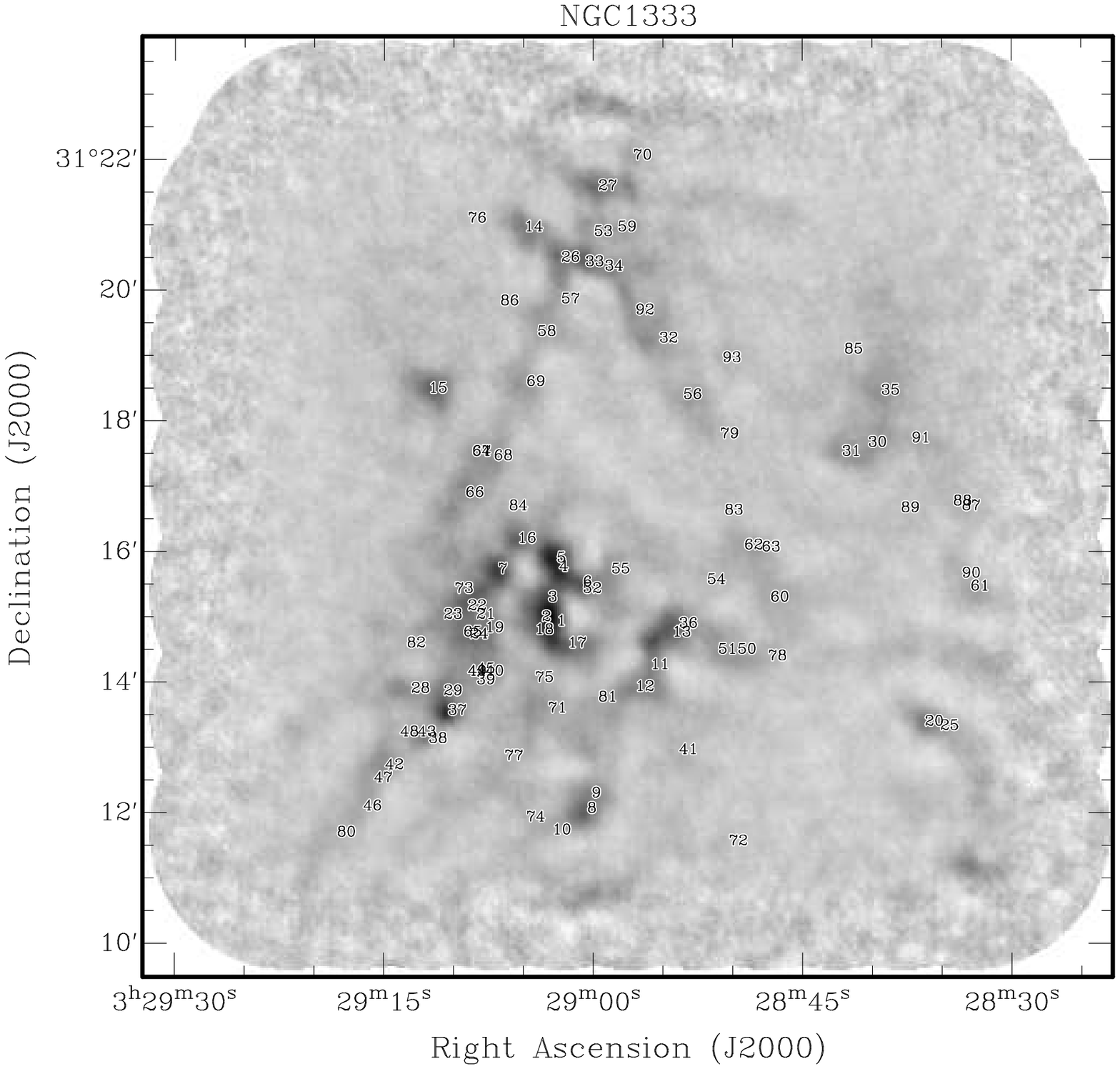}{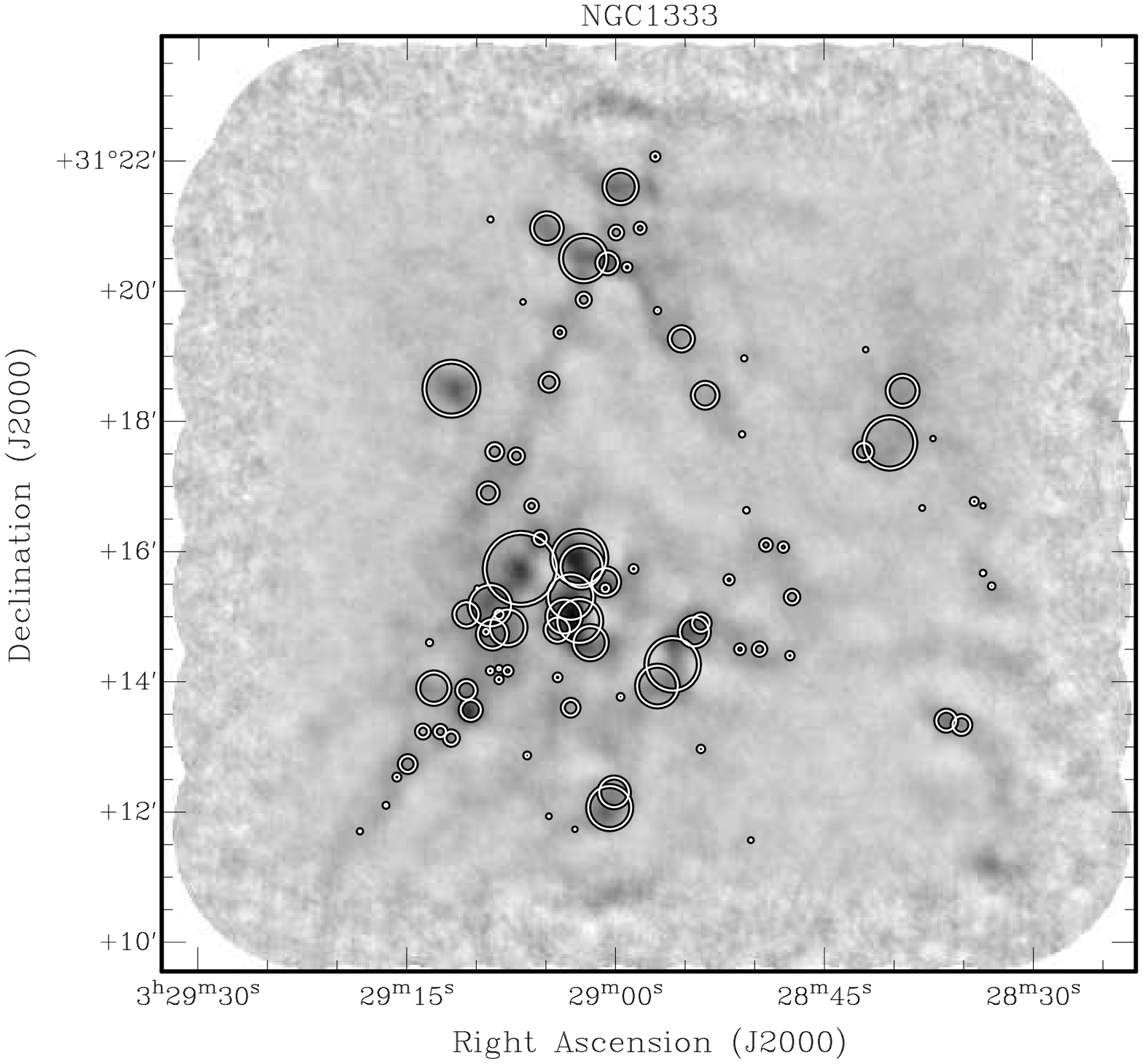}
\caption{Distribution of identified \nh~cores. The greyscale is the \nh~integrated intensity map.
Numbers on the left correspond to the positions of cores listed in Table \ref{n2hresults}.
The radius of the rings on the right is proportional to the ${\rm M}_{\rm LTE}$ for each core,
listed in Table \ref{n2hresults}. Note: weak cores are seen on the edges of the map,
which were not identified by {\sc clumpfind} as they lie outside the fully sampled region of
the BIMA observations. An example is located at 03 28 33.0 +31 11 12.}
\label{n2hcores}
\end{figure}

\begin{figure}
\plotone{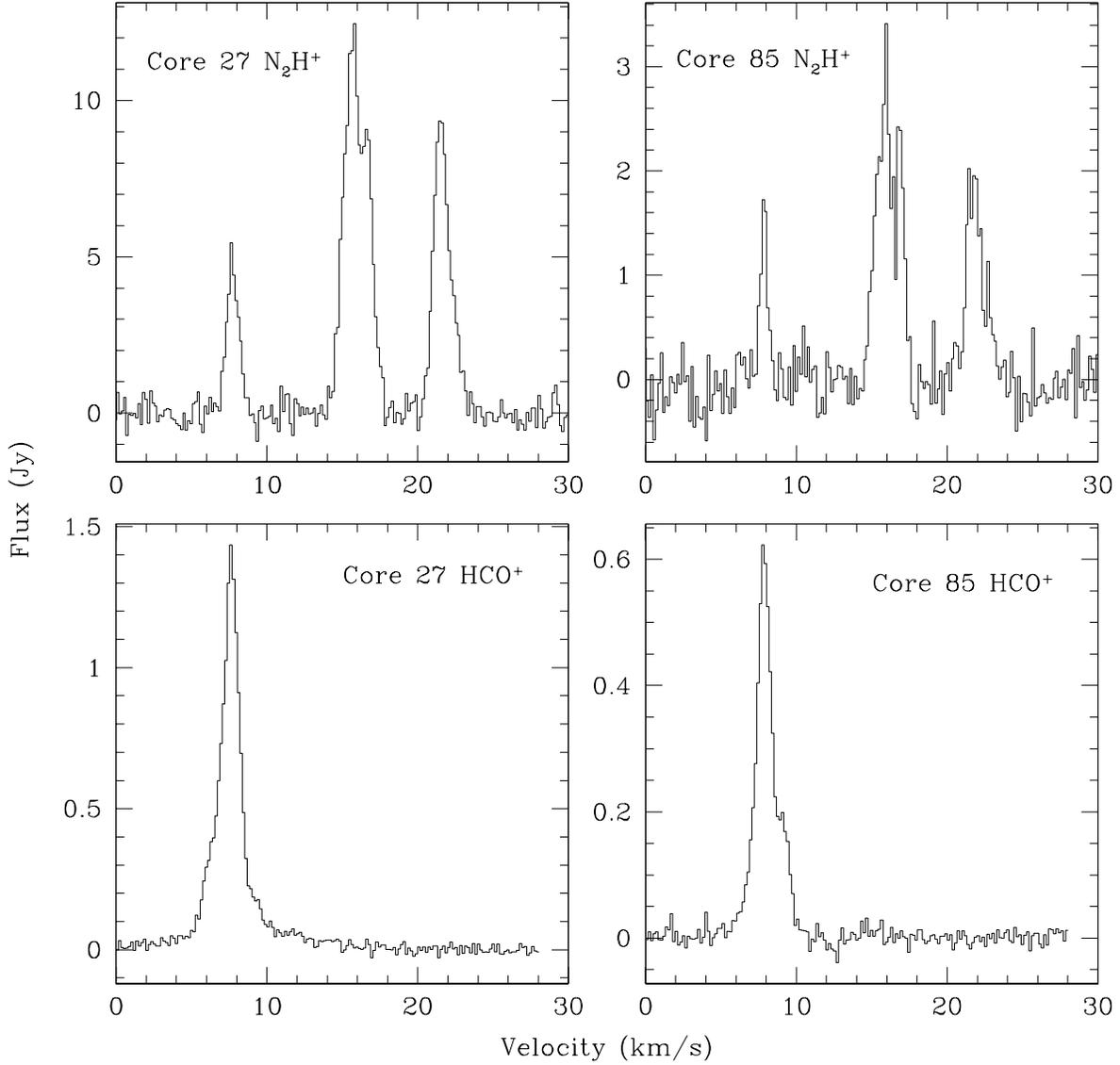}
\caption{N$_2$H$^+$ and HCO$^+$ spectra for two example cores. The left hand side shows N$_2$H$^+$ (upper)
and HCO$^+$ (lower) spectra for core 27, which has an LTE mass of 1\,${\rm M}_\odot$. The right hand side
shows N$_2$H$^+$ (upper) and HCO$^+$ (lower) spectra for core 85, which has an LTE mass of
0.05\,${\rm M}_\odot$. In both cases, the N$_2$H$^+$ emission is clearly detected, giving us confidence
that each identified core is real.}
\label{spectra}
\end{figure}

The details of the cores are shown in Table \ref{n2hresults} and their
relative positions are shown in Figure \ref{n2hcores}. As mentioned previously, there are five dust continuum 
cores to the NE (SK01) that do not exhibit any \nh~emission. Apart from these, nearly every SK01 continuum core
coincides with an \nh~core. The only exception is core 17 from SK01, located to the SW of IRAS\,6, which also does not
show any \nh~emission. Figure \ref{spectra} shows example \nh~and \hcop~spectra
for two cores (27 and 85), which are examples of high and low mass cores, respectively. It is clear that
the \nh~emission towards core 85 (the lower mass core) is well detected. As can be seen in Figure
\ref{n2hcores}, many of the cores are found in filamentary structures of high aspect ratio. These filaments
are seen extending to the
SE and SW of IRAS\,6 and to the SE of SSV\,13, through IRAS,4. The filaments appear to closely
(but not exactly) follow the dust distribution (SK01). We also calculate the core mass by two different
methods: the LTE approximation and by using the virial theorem. These are discussed in detail in the
following sections.

\subsubsection{LTE Mass}
We assume that the column density traced by the \nh~(1--0) transition in a core is given by \citep{williams94}
\begin{equation}
{\rm N} = \frac{3 \epsilon_0\,{\rm k}}{\pi^2 \nu \mu^2} \left(1-{\rm e}^{\frac{-{\rm h} \nu}{\rm k T}}\right)^{-1} \left(\frac{{\rm T}}{({\rm J}[{\rm T}]-{\rm J}[2.73])}\right) \int{\rm T}({\rm v})\;{\rm dv}
\label{eq1}
\end{equation}
where
\begin{equation}
{\rm J}[{\rm T}] = \frac{{\rm h} \nu}{{\rm k} \left({\rm e}^{\frac{{\rm h} \nu}{{\rm k} {\rm T}}} - 1\right)}
\label{eq2}
\end{equation}
where $\epsilon_0$ is the permittivity of free space; $\mu$ is the dipole moment for \nh~(1--0), which is
3.4\,Debye ($1.13 \times 10^{-29}$\,C\,m) and $\int{\rm T}({\rm v})\;{\rm dv}$ is the line integrated
intensity. We assume the temperature of the gas (T) to be 15\,K. We choose 15\,K because it is on the upper
limits for temperatures of isolated cores \citep{benson89} and we expect temperatures to be slightly
higher in clustered regions. However, we caution that uncertainties in temperature may lead to an error as large as
a factor of a few.
To convert this to a hydrogen column density, we assume the relative abundance of \nh~is $1.8 \times 10^{-10}$.
We find values quoted in the literature from $3 \times 10^{-11}$ \citep{bergin02} for the isolated core B68
to $5\times 10^{-9}$ for IRAS\,2 in NGC\,1333 \citep{jorgensen04a}, which both appear to be extreme cases.
Reasons for assuming this abundance are given in \S\ref{compsec}, but we note here that
our assumed relative abundance lies within the extremes and will be a representative value for most cases, although we might expect deviations
of up to an order of magnitude in extreme situations. Thus, the hydrogen column density is given by:
\begin{equation}
{\rm N}_{\rm H} =  9 \times \frac{\rm N}{1.8 \times 10^{-10}}
\label{eq3}
\end{equation}
The factor of 9.0 in the above equation is determined by measuring the average ratio of the flux of the
isolated \nh~hyperfine component to the total flux, and assumes the line is optically thin.
The LTE mass is then calculated using:
\begin{equation}
\left(\frac{{\rm M}_{\rm LTE}}{{\rm M}_\odot}\right) =
3.756 \times 10^{-32} \; \overline{m} \; \left(\frac{{\rm N}_{{\rm H}_2}}{{\rm m}^{-2}}\right)
\left(\frac{{\rm T}_{\rm av}}{\rm K}\right) \; \left(\frac{\rm d}{\rm pc}\right)^2 \;
\left(\frac{\rm pixsize}{\rm arcsec}\right)^2 \; \left(\frac{\Delta {\rm V}}{{\rm km}\,{\rm s}^{-1}}\right) \;
{\rm N}_{{\rm PIX}}
\label{eq4}
\end{equation}
where $\overline{\rm m}$ is the mean molecular weight of the gas, taken to be 2.3; ${\rm N}_{{\rm H}_2}$ is
taken from Equation \ref{eq3} above and assumes the gas is molecular; ${\rm T}_{\rm av}$ is the mean
brightness temperature of each pixel of
the core identified by {\sc clumpfind}; d is the distance to NGC\,1333, which is taken to be 300\,pc \citep{dezeeuw99,belikov02};
pixsize is the size of the data cube pixels in arcseconds, which is 2\arcsec~for our data; $\Delta {\rm V}$
is the velocity width of each channel, $0.157\,{\rm km}\,{\rm s}^{-1}$; ${\rm N}_{\rm PIX}$ is the number of
pixels that are identified with the core.

The uncertainty in the calibration of our flux density scale will contribute an error to the derived LTE masses.
We estimate this error to be no greater than 20\%. Our assumed distance of 300\,pc comes from HIPPARCOS
measurements, however we note distances in the literature ranging from 220 (SK01) to 350\,pc
\citep{difrancesco01}. Thus masses and sizes
will need adjusting when comparing to work using different distances.
We also anticipate that uncertainty in the temperature of
the gas (we assume 15\,K) will also contribute an error which may be as large as a factor of a few. We believe
the error associated with the uncertainty in the relative abundance of \nh~to be the largest error in the LTE
mass calculation.

\subsubsection{Virial Mass}
We can also use the \nh~data to calculate virial masses for the cores using:
\begin{equation}
{\rm M}_{\rm VIR} = 210 \times \delta {\rm V}^2 \; {\rm r}
\label{eq5}
\end{equation}
where $\delta {\rm V}$ is the line FWHM and r is the effective radius of the core (column 5 of Table
\ref{n2hresults})\citep{caselli02}. Equation \ref{eq5} assumes a uniform density profile. If we assume a density profile that
falls off as ${\rm r}^{-2}$, then the constant 210 in Equation \ref{eq5} changes to 126 \citep{williams94}.
Thus, our calculation of the virial mass of the cores may decrease by a factor of two. We also note
that the virial mass ignores magnetic fields, as well as the effects of an external pressure, which may have a
significant effect on the virial mass in a region such as NGC\,1333, due to mechanical pressure from outflows
impinging on the cores. Furthermore, outflows are likely to be hot, which also may increase the pressure on the cores.

\subsubsection{Comparison of ${\rm M}_{\rm LTE}$ and ${\rm M}_{\rm VIR}$}
\label{compsec}
\begin{figure}
\plotone{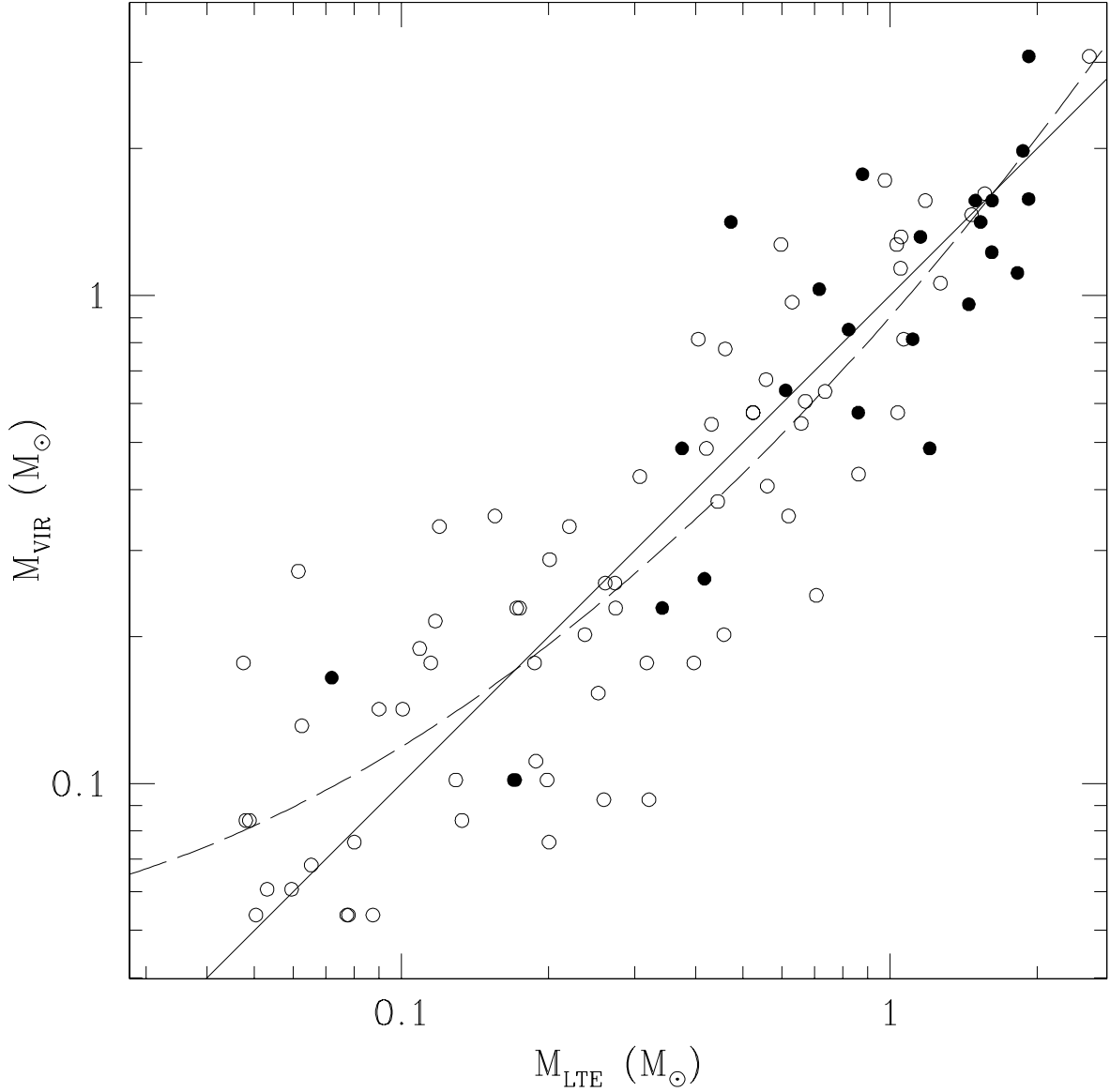}
\caption{Distribution of core LTE and virial masses. Filled circles represent cores associated with stars, as
detected by {\it Spitzer}, and open circles represent cores not associated with stars.
The solid diagonal line represents equality between ${\rm M}_{\rm LTE}$
and ${\rm M}_{\rm VIR}$, assuming an \nh~relative abundance of 1.8 $\times 10^{-10}$.
The dashed curve represents a best fit to the data of the form y = a + bx + c${\rm x}^2$, where x
is Log(${\rm M}_{\rm LTE}$), y is Log(${\rm M}_{\rm VIR}$), a is 0.071, b is 1.38 and c is 0.26.}
\label{masscomp}
\end{figure}

The LTE mass is a measure of the amount of matter located within a core, whereas the virial mass is a measure of
amount of matter required to keep a core bound. Therefore, we can compare the two masses to evaluate whether
a core has enough self gravity to stay bound or not. If a core is bound, then the LTE mass should be greater
than, or equal to, the virial mass. The distribution of virial masses with respect to
LTE masses for the same core is shown in Figure \ref{masscomp}. We have chosen a value of the \nh~relative abundance
($1.8 \times 10^{-10}$) such that the points in Figure \ref{masscomp} are distributed evenly about the line
of equality (${\rm M}_{\rm LTE} = {\rm M}_{\rm VIR}$), shown by the solid line.
If we assume a different relative abundance then this will have the
effect of sliding the diagonal line of equality along the horizontal axis in Figure \ref{masscomp}.
Because of the uncertainty in the abundance, we cannot be sure whether individual cores are bound or unbound
based solely on a comparison of their LTE and virial masses.

M$_{\rm VIR}$ and M$_{\rm LTE}$ are correlated with a correlation coefficient of 0.88 over a range in
M$_{\rm LTE}$ of $\sim$40, despite the individual uncertainties in M$_{\rm VIR}$ and M$_{\rm LTE}$. Such a
correlation is unlikely to be a coincidence, suggesting that most of the cores are self-gravitating, at both
high and low ends of the mass range.
If proportionately more of the lowest mass cores were unbound, compared to the
highest mass cores, then we would expect to see the distribution of points in Figure \ref{masscomp}
to turn up towards the virial mass axis, in the lower left (lowest mass) corner. However, there are
approximately equal numbers of
points above (19) and below (18) the solid line with ${\rm M}_{\rm LTE} < 0.3$, as compared to
approximately equal numbers of points above (29) and below (27) the solid line with
${\rm M}_{\rm LTE} > 0.3$. The curved line in Figure \ref{masscomp} is a second order polynomial
fit to the data points, given in Equation \ref{eqn6}:
\begin{equation}
{\rm Log}({\rm M}_{\rm VIR}) = 0.071 + 1.38\,{\rm Log}({\rm M}_{\rm LTE}) + 0.26\,({\rm Log}({\rm M}_{\rm LTE}))^2
\label{eqn6}
\end{equation}
Whilst the polynomial fit in Figure \ref{masscomp} does show a small turn up at the lower mass end,
the turn up is small compared to the scatter in the data points. Therefore we do not consider it significant.
Thus, we have no evidence for such a turning up of the distribution in our data. We conclude that many
of the lower mass cores are likely to be self-gravitating, and are not transient objects.

Some of the higher mass cores can be expected to be bound, since they are associated with protostars, such
as core 37 (IRAS\,4A), core 38 (IRAS\,4B), core 36 (IRAS\,2C) and core 15 (IRAS\,7).
Below we detail some of the properties of these cores.

{\bf Core 37 (IRAS\,4A)}. We calculate an LTE mass of $0.7\,{\rm M}_\odot$ and a virial mass of
$1.0\,{\rm M}_\odot$. This might suggest that the core is unbound, but we know this is not the case
because IRAS\,4A contains at least two protostars \citep{lay95}: an alternate explanation is needed.
IRAS\,4A was observed by \citet{difrancesco01} in \nh~at a higher resolution
than our observations. They calculate the LTE mass of IRAS\,4A to be $0.73 {\rm M}_\odot$, assuming the distance
to NGC\,1333 is 350\,pc and X(\nh) = $6.0 \times 10^{-11}$. Converting this mass using our preferred distance
(300\,pc) and \nh~relative abundance ($1.8 \times 10^{-10}$), this is equivalent to $0.18 {\rm M}_\odot$,
a factor of 4 lower than our derived value. This is partly because their higher resolution observations filtered
out some of the extended emission. By combining the BIMA data with the FCRAO data, we have minimised this problem.
The relative abundance of \nh~makes a significant difference to the derived LTE mass. We can check our derived
relative abundance by comparing LTE mass to that derived from dust continuum measurements by SK01. SK01
determine the mass of IRAS\,4A to be 1.5\,${\rm M}_\odot$, assuming a distance of 220\,pc. To reproduce
our LTE mass, we require a relative abundance of $5 \times 10^{-11}$, similar to that derived by
\citet{difrancesco01}. Therefore, we agree with the finding of \citet{difrancesco01} that there is an unusually
low relative abundance of \nh~at IRAS\,4A. We see the effect of this lowered abundance when comparing the relative
intensities of \nh~and 850$\mu$m continuum emission at IRAS\,4A in Figure \ref{n2hp}. The continuum contours
strongly peak at the position of IRAS\,4A whilst the \nh~emission is only a weak peak at this position.

A possible explanation for the low abundance is that IRAS\,4A is the source of an outflow, which is traced
by \hcop.
The \hcop, and CO, will be enhanced in the gas phase by the shocks from the outflow. Since both \hcop~and
CO destroy \nh~\citep{jorgensen04b}, we expect to find a lower \nh~abundance here. An alternate explanation
is that IRAS\,4A has undergone thermal processing of its \nh, thereby reducing the \nh~abundance.

{\bf Core 38 (IRAS\,4B)}. We calculate an LTE mass of 0.5\,${\rm M}_\odot$ and a virial mass of
$1.4\,{\rm M}_\odot$. Again, this might imply the core is unbound, but we know this is not the case \citep{lay95}.
\citet{difrancesco01} have also observed this core at higher resolution and find an
LTE mass of 0.41\,${\rm M}_\odot$, using the same assumptions as for IRAS\,4A. Here, we assume that again
there is a significant amount (a factor of 5) of missing \nh~flux in the data of \citet{difrancesco01}.
Comparing our LTE calculations to the dust continuum mass derived by SK01, we can estimate the relative
\nh~abundance. SK01 derive a combined mass of 0.63\,${\rm M}_\odot$ for IRAS\,4B west and east. We combine the
masses for the west and east components here because our resolution was not sufficiently high enough
for {\sc clumpfind} to separate \nh~emission surrounding the two protostars. The \nh~abundance is then estimated to be
$1 \times 10^{-10}$, close to assumed abundance of $1.8 \times 10^{-10}$.

{\bf Core 36 (IRAS\,2C)}. We calculate an LTE mass of 0.6\,${\rm M}_\odot$ and a virial mass of
0.4\,${\rm M}_\odot$. IRAS\,2C has been identified as one of the youngest sources in NGC\,1333 and is
most likely a Class 0 source (SK01). SK01 also use their dust continuum observations to derive a mass
of 0.05\,${\rm M}_\odot$. In this case, it appears there is an \nh~abundance enhancement of about 6 over
the assumed average value.

{\bf Core 15 (IRAS\,7)}. We calculate an LTE mass of 2\,${\rm M}_\odot$ and a virial mass of
1.6\,${\rm M}_\odot$. SK01 calculate a mass of 0.16\,${\rm M}_\odot$ for the combined contributions of
SM\,1 and SM\,2. Again, it appears the abundance may be enhanced by a factor of a few over the average
value.

\subsection{Association of cores with stellar sources}
Recent observations of NGC\,1333 with the {\it Spitzer Space Telescope} \citep{gutermuth06} allow us to
compare the incidence of our \nh~cores with infrared stars. All the stars we correlate with \nh~cores
are known to be pre-main sequence stars within NGC\,1333, by nature of their infrared colors, using the bands
at 3.6, 4.5 and 5.8$\mu$m, which identifies Class I, II and III sources. In addition to this, we used 24 and 70$\mu$m
{\it Spitzer} images to associate cores with Class 0 sources. We note that all previously known Class 0 sources in the
field of view are easily detected at these longer wavelengths.
The infrared observations have a sensitivity limit
of around 0.05\,${\rm M}_\odot$, which is similar to the mass limit of our detected cores.
We consider a star associated
with an \nh~core if any part of the \nh~core, as defined by {\sc clumpfind}, overlaps with the peak of the
infrared emission.

Using the same association critera, we have used a monte carlo simulation to test how many chance
alignments of \nh~cores and infrared sources we might expect. To do this, we assume an average \nh~core size of 0.015\,pc
and randomly place 93 simulated cores in the field of view and count the number of chance alignments. This process is
repeated to derive a distribution of expected chance alignments. From these simulations, we expect 3.5 chance alignments
of \nh~cores with infrared stars. This low number of chance alignments is unlikely to affect our results discussed below.

\begin{figure}
\plotone{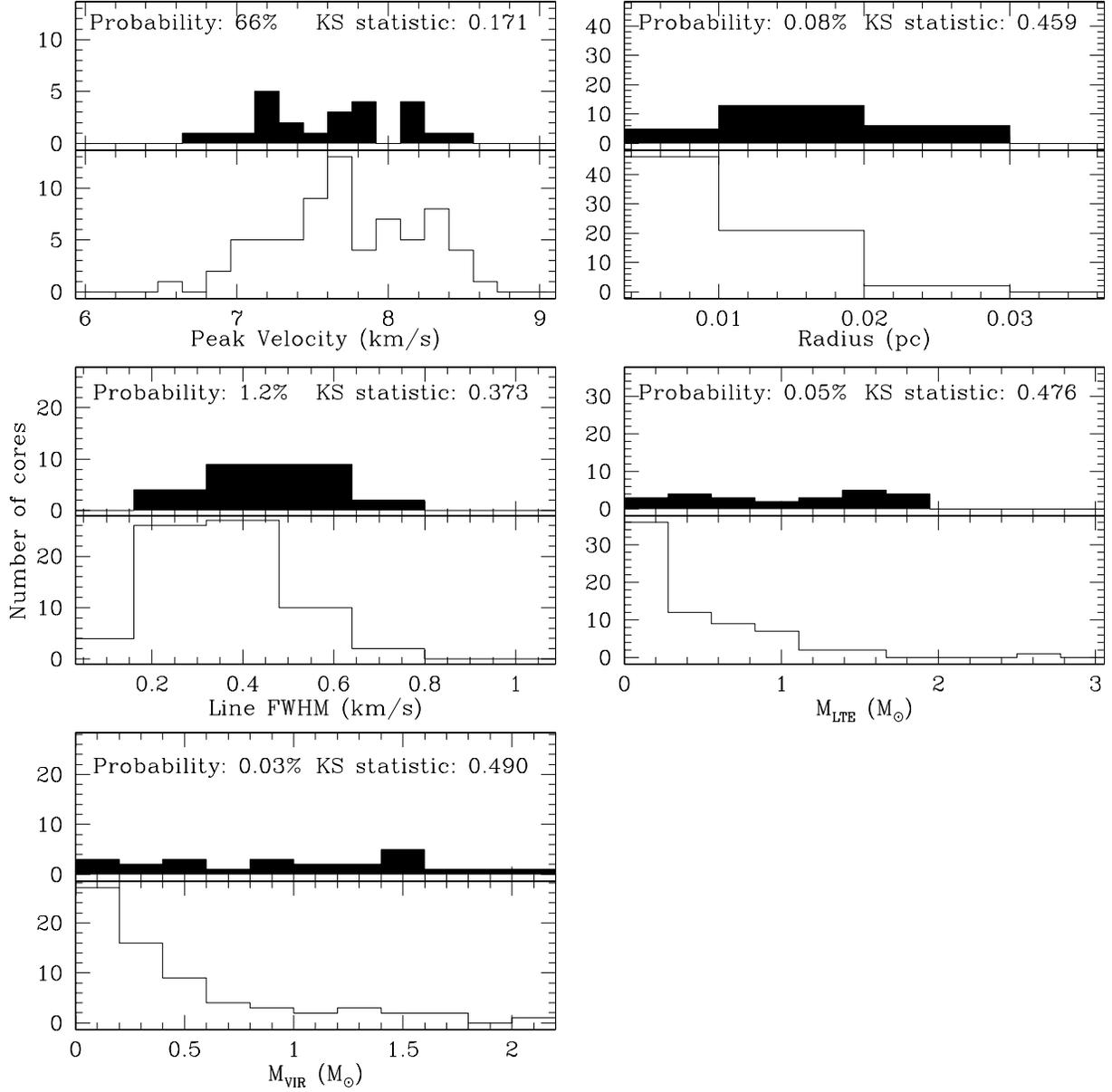}
\caption{Distributions of \nh~core parameters depending on whether they are associated with stars
(solid histograms) or not (open histograms). For each pair of distributions, a Kolmogorov-Smirnov
(KS) test is applied and the resulting KS statistic and probability is given. This is the
probability that the two distributions are drawn from the same sample. Significant differences
are seen in all distributions except peak velocity.}
\label{starnostar}
\end{figure}

\nh~cores with associated stars are shown in Figure \ref{masscomp}. It is clear from this figure that
\nh~cores associated with stars tend to have higher masses than those without. This can also be seen in Figure
\ref{starnostar}, as well as the distributions of other attributes of the \nh~cores. Both the distributions
of \nh~core LTE and virial masses are extremely unlikely to come from the same population.
Figure \ref{starnostar} also shows that cores without associated stars are generally smaller than those with stars.
Implications of this are discussed further in \S\ref{massdist}.

The \nh~core line FWHM is generally smaller for cores not associated with stars. This is what would be expected
if the stars are active in creating local turbulence through their stellar winds and outflows. Such winds and
outflows should stir up the immediate environment of surrounding gas, increasing the measured line FWHM.
The tendency of cores with associated stars to have larger line widths has been seen in ammonia cores
\citep{benson89,jijina99}.

%
%
%
\subsection{Mass distribution of the cores}
\label{massdist}
Recent simulations by \citet{bonnell01a} and \citet{bate05} indicate that the initial mass function (IMF) is created
at an advanced stage of evolution. Initially all cores start with the same mass, which is determined
by the opacity limit for fragmentation. Stars build up their mass through accretion. The final
mass of the star is then determined by stochastic ejection of the stars from the cluster due
to dynamical interactions. However, previous observations of pre-stellar dust cores in the rho Ophiuchi
main cloud \citep{motte98,johnstone00} show that the mass distribution
of some of the pre-stellar cores resembles the IMF.
This suggests that the IMF may be locked in at the pre-stellar stage, in contradiction to the simulations.
The previous observations by \citet{motte98} and \citet{johnstone00} were continuum measurements and so
no linewidth information for the cores is known. It is therefore possible that the pre-stellar cores are
transient phenomena and may not form stars, as suggested by \citet{ballesteros02}.
Our \nh~data have shown that in NGC\,1333
at least, the \nh~cores appear to be close to virialised and therefore a significant proportion
are likely to be bound and are therefore more likely to form stars.

\begin{figure}
\plotone{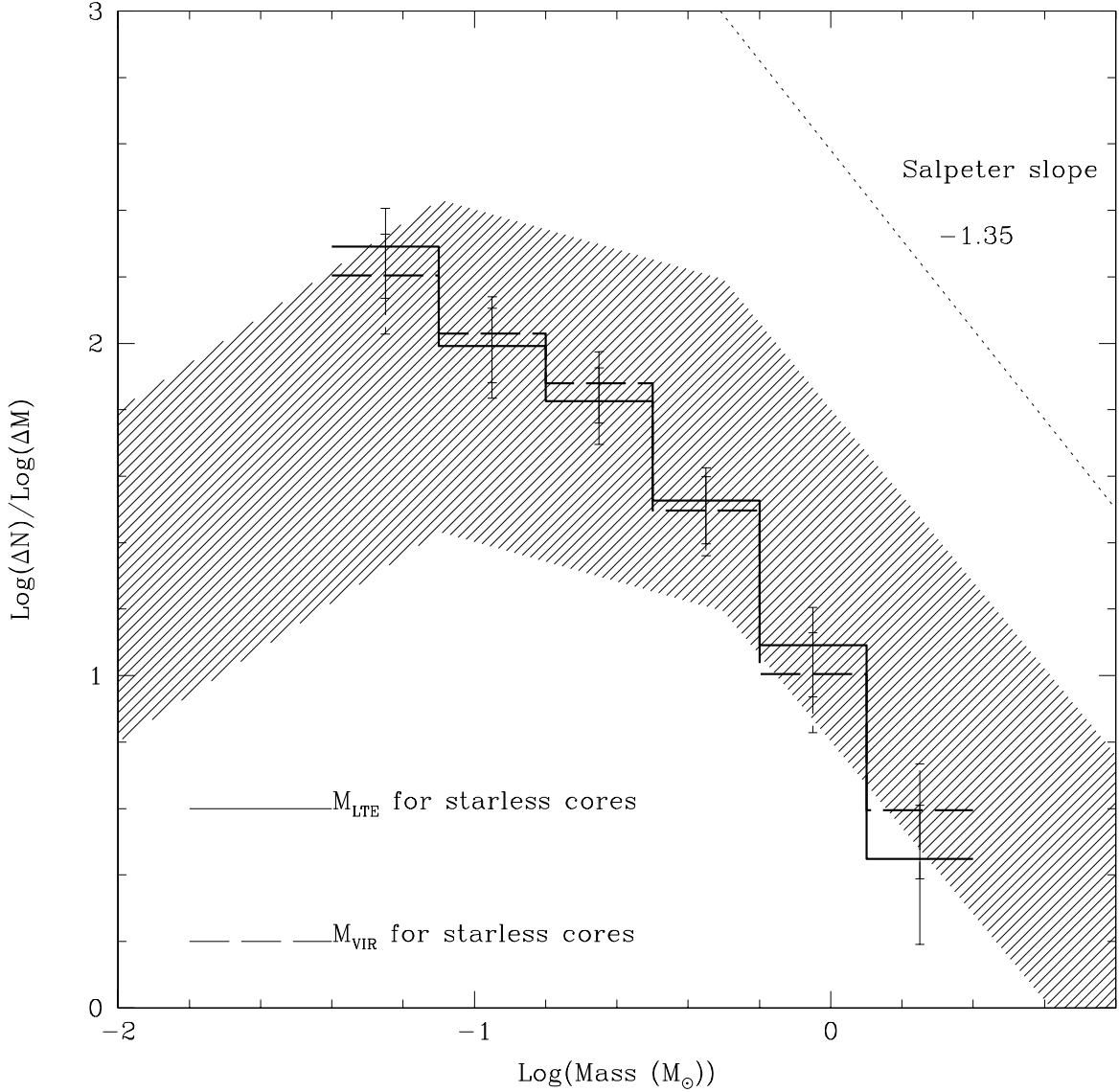}
\caption{Distribution of starless core LTE and virial masses derived from the N$_2$H$^+$ data.
The solid line histogram shows the LTE masses and the
dashed line histogram shows the virial masses, with error bars representing $\sqrt{\rm N}$ statistics.
The LTE masses were determined, assuming a relative abundance of 1.8 $\times 10^{-10}$ for \nh.
The shaded area represents the average field star IMF, with uncertainty, as determined by \citet{kroupa02}.
The dotted diagonal line represents the Salpeter IMF slope \citep{salpeter55}. Both the starless
core LTE and virial masses are consistent with the field star IMF, given the error bars represented
by the shaded area.}
\label{mass_distn}
\end{figure}

In Figure \ref{mass_distn}, we show the distribution of core LTE and virial masses. The LTE and virial mass
distributions are very similar. Furthermore, we find these distributions similar to the average field star IMF,
although the LTE and virial mass distributions do not show a decrease for lower mass cores, as is seen in the
IMF.
%
%
Such similarities between the stellar IMF and the the mass function of dust cores has also been found
by \citet{motte98}, \citet{johnstone00} and \citet{stanke06}, who suggested this is evidence that the
IMF may be locked in at this
early stage of star formation. A problem with comparing the mass function of dust cores and stars
is that it is not clear whether the dust cores are bound or not -- no information on their internal motions
is known. The \nh~data allow us to overcome this problem. If the \nh~cores are bound regardless of mass, this
suggests that such emission does not trace transient structures formed out of turbulent fluctuations.
Therefore, this implies the entire mass distribution consists of objects that may potentially form stars.

It is interesting to note that in Figure \ref{starnostar}, we see a flat distribution of core masses
(both LTE and Virial) for those cores associated with stars. This flat distribution is clearly unlike
the field star IMF. There appear to be too few low mass cores to form a mass function similar to
the field star IMF. Because both the \nh~and infrared observations have
a sensitivity limit of around 0.05\,${\rm M}_\odot$, we do not believe this is an effect of the
sensitivity limits.

As a core forms a star, it loses gas mass. If lower mass cores lose a bigger proportion of their mass
than do higher mass cores, the lower mass cores move left in the distribution shown in Figure \ref{starnostar}
more than the higher mass cores, making the mass function distribution shallower. But if more of the high mass
cores make multiple stars than do low mass cores, the mass function should look steeper. These are two competing effects
that may well affect the distribution we see. Future research should be undertaken to identify the
cause of this apparent flattening of the mass function.

%

\subsection{Clustering}
\begin{figure}
\plotone{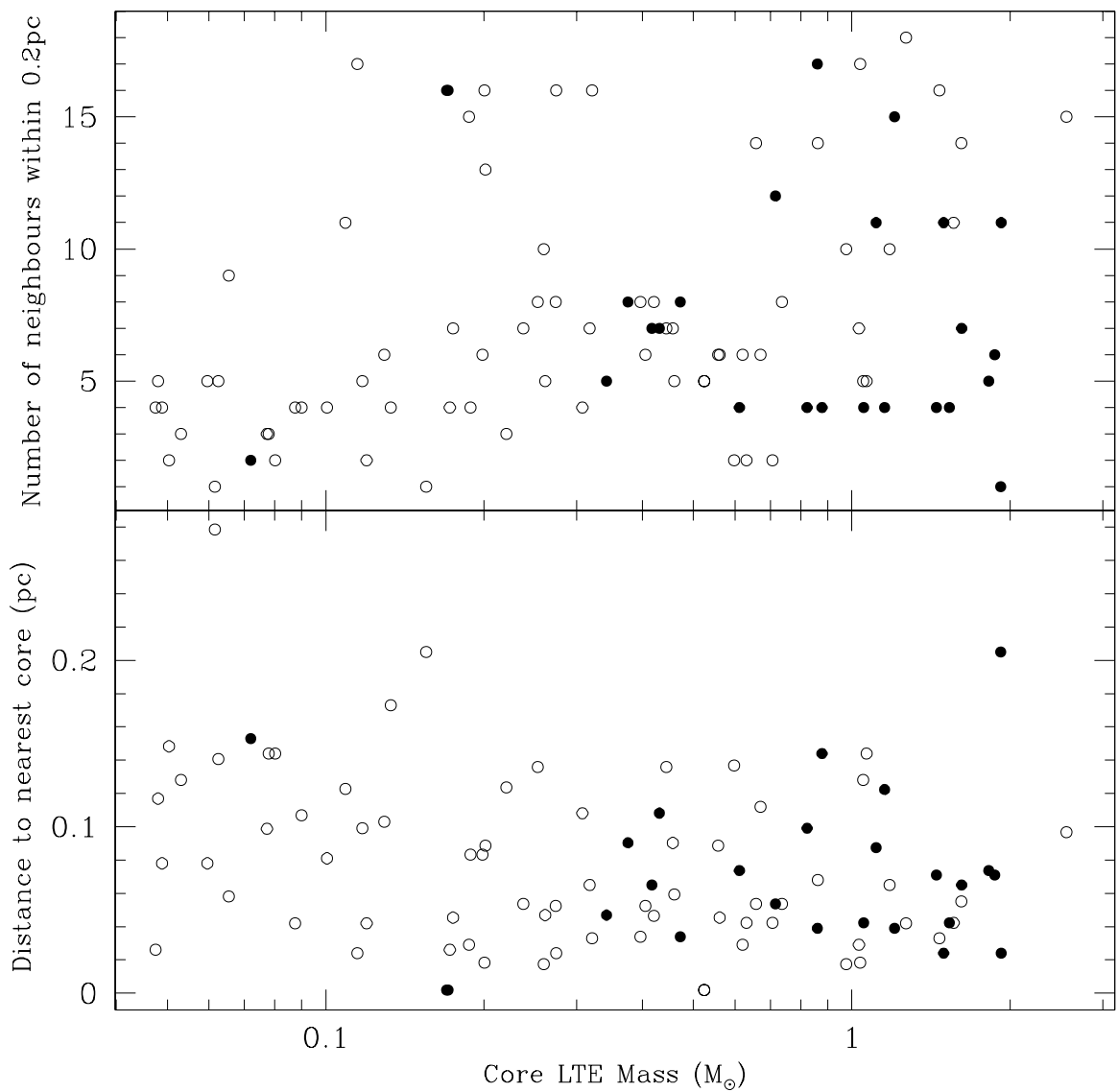}
\caption{Distribution of clustering properties for \nh~cores. The upper plot shows the number of cores
within 0.2\,pc of each core. The lower plot shows the distance to the nearest core. Filled in circles
represent those cores that are associated with stars. Open circles represent cores that are not
associated with stars. Both plots show that there is no correlation between the clustering properties of
cores with high or low mass.}
\label{clustering}
\end{figure}

NGC\,1333 is known as a region of clustered star formation \citep{aspin94,lada96}. As can be seen from
Figure \ref{n2hcores}, the \nh~cores also appear to cluster, showing a concentration in the
IRAS\,4/IRAS\,2/SSV\,13 region, with a particular concentration of cores around SSV\,13. In Figure
\ref{clustering}, we show a comparison of clustering properties of \nh~cores for varying
LTE masses. For each core, the upper plot shows the number of cores that are within a radius of 0.2\,pc,
whilst the lower plot shows how far away the nearest core is. Both distributions are a measure of the clustering
properties of the cores throughout their mass spectrum. As can be seen, there is no clear correlation
between either of the distributions. However, it appears that only a small proportion of cores with
M$<0.2$\,M$_\odot$ have more than four neighbours (43\%), whereas proportionately more of the higher mass cores
have more than four neighbours(79\%), But this trend is weak and does not appear for other choices mass
cutoff. This implies that there is little significant increase in the likelihood
of finding a more massive core in a clustered region, than a lower mass core. Furthermore, when cores
are separated into those that are associated with stars (filled in circles) and those that are not (open
circles), there is no significant difference between the two populations.

\subsection{Motions of the cores}

An important difference between isolated and clustered star formation is the effects of turbulence, which
are expected to be larger in the clustered environment. Di Francesco, Andr\'{e} \& Myers (2004) have studied
the relative motions of cores in the Ophiuchus A cluster region.
The Oph A observations have a higher spatial resolution ($10\arcsec \times 6\arcsec$, corresponding
to $1250 \times 750$ AU) than the NGC\,1333 observations ($10\arcsec \times 10\arcsec$, corresponding to 3000\,AU)
but cover a smaller area ($\sim$0.1\,pc for Oph A and $\sim$1.0\,pc for NGC\,1333).
\citet{difran04} found little dispersion
between the line-of-sight velocities of the cores they detected within Oph\,A with an rms dispersion of
only 0.17\,km\,s$^{-1}$. For NGC\,1333, we find a mean line-of-sight velocity of 7.66\,km\,s$^{-1}$ and an rms
dispersion of 0.46\,km\,s$^{-1}$. This is larger than that found for Oph\,A. Examination of the \nh~data cube
suggests that this is because there appear to be two regions of \nh~emission along the same line of sight in
the vicinity of SSV\,13. These are traced by cores 4, 5, 6, 7, 16 and 84 which all have velocities greater than
8.1\,km\,s$^{-1}$, and cores 1, 2, 3, 17, 18, 52 and 55 which all have velocities less than 7.8\,km\,s$^{-1}$.
For each velocity component, we find the rms dispersion is 0.14 and 0.34\,km\,s$^{-1}$ for cores greater than
8.1 and less than 7.8\,km\,s$^{-1}$, respectively. Thus,
it appears the velocity dispersion is larger in NGC\,1333 than in Oph A, even taking into account the multiple
velocity components along the line of sight in NGC\,1333.
For the \nh~core line widths, we find an average value of
0.38\,km\,s$^{-1}$ and an rms dispersion of 0.14\,km\,s$^{-1}$. These numbers are very similar to those found
by \citet{difran04} in Oph\,A: 0.39\,km\,s$^{-1}$ and 0.11\,km\,s$^{-1}$, respectively. So it appears that
individual core kinematic properties are similar in Oph A and NGC\,1333, but there are larger relative motions
of the cores in NGC\,1333. However, this may be an effect for measuring a much larger area in NGC\,1333. In order
to compare the two clusters in more detail, more cores need to be identified on a similar size scale to the Oph A
observations.

We can investigate the motions of cores within the filamentary structures to the SE of SSE\,13 and to the 
SW and SE of IRAS\,6. If we assume that the filamentary structures are true filaments; ie. that they will
also appear as filaments when observed orthogonal to the current viewing angle, then we can estimate how long
it will take for cores to move the length of a filament, based on the observed velocity dispersion. This will
tell us the typical timescale we might expect the filaments to survive. We find velocity dispersions of cores
within the filament to the SE of SSV\,13 (0.6\,pc long), and to the SW (0.5\,pc long) and SE (0.5\,pc long) of
IRAS\,6 to be 0.23, 0.27 and 0.24\,km\,s$^{-1}$, respectively. Taking an average of 0.25\,km\,s$^{-1}$,
and assuming cores must
travel 0.5\,pc for a filament to dissipate, we calculate the filaments may survive for $2 \times 10^6$ years.
Such a long time implies these filaments may well be stable features.

Previously, Walsh, Myers \& Burton (2005) have investigated the relative motion of star forming cores with respect
to their surrounding envelopes. This was done by observing \nh (1--0) which traces the high density core
motions and $^{13}$CO (1--0) and C$^{18}$O (1--0) which trace the low density surroundings. They found little
evidence for any ballistic relative motions. This is in contradiction to previous theoretical work
\citep{bonnell97,bonnell01b} which suggested that a star's final mass was determined by its motion through its natal cloud.
The observations of \citet{walsh05} were focused mainly towards regions of isolated star
formation, with a very small number of regions close to clustered environments.
Their results were clear for the case of the
isolated regions of star formation, but inconclusive for the clustered regions. With these new data, we are able to 
look for relative motions in NGC\,1333 and thus extend the work to include a clustered region of star formation.

\begin{figure}
\plotone{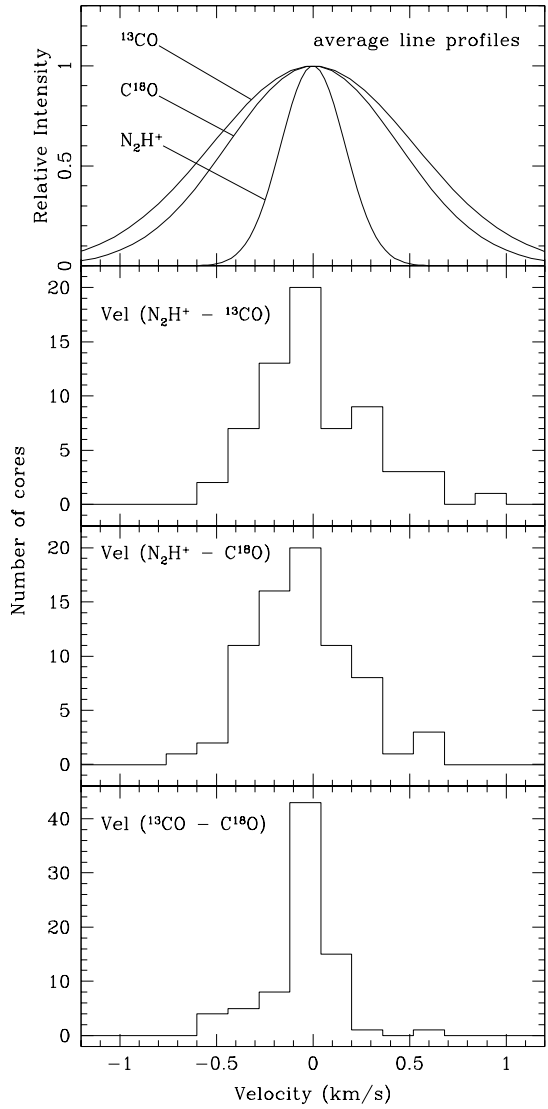}
\caption{Line center velocity differences for \nh~, $^{13}$CO and C$^{18}$O are shown on the lower
three plots. The top plot shows simulated line profiles of the three transitions, based on their
average line profiles.}
\label{veldiffs}
\end{figure}

We use the line center velocities of the \nh~cores given in Table \ref{n2hresults} as our high density tracer
and the $^{13}$CO (1--0) and C$^{18}$O (1--0) FCRAO maps of \citet{ridge03} covering NGC\,1333 as the low
density tracers.
We determine the difference in line center velocities by using the \nh~core velocities
given in Table \ref{n2hresults} and fitting a gaussian to the $^{13}$CO and C$^{18}$O line profiles of the
spectrum that coincides with the peak of the \nh~core. Not all \nh~cores
could be included in the analysis. This is because some of the CO spectra were too confused to reliably determine a relevant
line center velocity. Figure \ref{veldiffs} shows the difference in the line center
velocities, which can be compared to Figure 1 of \citet{walsh05} for the case of isolated cores.

The top of Figure \ref{veldiffs} shows simulated line profiles for the average line width of \nh, $^{13}$CO
and C$^{18}$O. As was previously discussed by \citet{walsh05}, if we expect there to be no motions of the high density
cores with respect to their surroundings, then we would expect to see line center velocity differences distributed like
the line profile for the \nh~transition. If there are significant motions of the core, then we would expect to see
broader distributions resembling the CO line profiles. Figure \ref{veldiffs} shows that line center velocity differences
have a distribution somewhere between these two extremes: The rms difference between \nh~and $^{13}$CO
is 0.57\,km\,s$^{-1}$ and between \nh~and C$^{18}$O is 0.53\,km\,s$^{-1}$. These numbers can be compared to the
average $^{13}$CO FWHM of 1.18\,km\,s$^{-1}$, average C$^{18}$O FWHM of 1.00\,km\,s$^{-1}$ and average \nh~FWHM of
0.38\,km\,s$^{-1}$.
Unfortunately, our results may suffer from some errors that were avoided by \citet{walsh05}. This
is because we are using different telescopes, and observations were made under different conditions. One
of the greatest source of errors is that our BIMA \nh~data are of much higher spatial resolution than the FCRAO
CO data. Thus it is likely that the CO data trace somewhat different columns, compared to
the \nh~data. This will give an apparent increase in the differences of measured line center velocities between
\nh~and the CO isotopes. This has the effect
of rendering our results uncertain as to whether we can tell if there are relative motions of the cores, although
we note that it seems to be consistent with the cores not moving ballistically.
In order to obtain a clear answer, it will be necessary to obtain CO data with the same resolution as the \nh~data, which
is beyond the scope of this work.


\section{Summary and Conclusions}

We have used both the BIMA and FCRAO radiotelescopes to observe HCO$^+$\,(1--0) and \nh\,(1--0)
in the clustered star forming region NGC\,1333. Comparison with submm continuum images shows
that the \nh~emission very closely follows the dust continuum, showing it is an excellent
tracer of quiescent gas. HCO$^+$, on the other hand, shows overall similarities with dust continuum,
but appears more extended and also traces outflowing gas.

We have identified 93 \nh~cores in the field of view, with masses between approximately 0.05\,${\rm M}_\odot$
and 2.5\,${\rm M}_\odot$. Whilst uncertainties in the relative abundance of \nh~do not allow us to
conclude whether or not individual cores are bound, the trend of core properties suggests that most
will be bound. Furthermore, there is no up-turn of the distribution of core Virial-LTE masses,
which would suggest that the lowest mass cores are not bound and therefore transient phenomena.
Our conclusion is that even these low mass cores are likely to remain bound and eventually form
stars, or brown dwarfs. If the \nh~cores are bound regardless of mass, this suggests that such
emission does not trace transient structures formed out of turbulent fluctuations.
Therefore, this implies the entire mass distribution consists of objects that can potentially form stars.

The mass function of the \nh~cores is consistent with the field star IMF, suggesting that
the distribution of masses may be locked in at the pre-stellar phase, in contradiction
to theories of mass accumulation by competitive accretion. We find that the mass function of core associated
with stars is flat, and unlike the IMF. This may be because lower
mass cores lose more of their mass when forming a star, however, further research is required
to fully understand this.

We find the relative motions of cores within NGC\,1333 to have an rms dispersion of 0.46\,km\,s$^{-1}$,
larger than that found for Oph A by \citet{difran04}, but the line widths of the cores
(average is 0.38\,km\,s$^{-1}$) is similar to those found in Oph A. We may be seeing an intrinsic
increase in relative motions of cores in NGC\,1333, but cannot discount the possibility that the larger
motions are an artifact of the large size scale probed in NGC\,1333.

We have compared the line center velocities of the low density gas tracers $^{13}$CO and C$^{18}$O
to those of \nh~in an attempt to look for relative motions of the high density \nh~cores with
respect to their low density surroundings, as predicted by some theories of clustered star
formation. The relative motions have a dispersion of about 0.55\,km\,s$^{-1}$, which is smaller
than expected if the cores were moving ballistically ($\sim 1.1$\,km\,s$^{-1}$), but larger
than the expected dispersion if the cores were not moving ($\sim 0.38$\,km\,s$^{-1}$). We note
that the dispersion of core relative velocities may be artificially increased due to differences
in beam sizes
between the telescopes used to make the measurements. Thus, the data are consistent with no
ballistic motions of the high density cores, but this conclusion needs to be tested more rigorously.
We suggest further observations of \nh~and a CO transition may be
able to answer this question decisively if the same telescope is used for observations of both
transitions, and care is taken to minimise any systematic errors.

\acknowledgments
We acknowledge the thoughtful comments by an anonymous referee that have greatly improved the quality of this paper.
The BIMA array is operated with support from the National Science Foundation under grants
AST 99-81308 to the University of California, Berkeley, AST 99-81363 to the University of
Illinois, and AST 99-81289 to the University of Maryland.
The Five College Radio Astronomy Observatory is supported by NSF grant AST 02-28993.
This research has made use of NASA's Astrophysics Data System Bibliographic Services.

\clearpage

\begin{deluxetable}{cccccccc}
\tabletypesize{\scriptsize}
\tablecaption{\nh~core properties}
\tablewidth{0pt}
\tablehead{
  \colhead{Core}&\multicolumn{2}{c}{Position (J2000)}&\colhead{Peak}&\colhead{Radius}&\colhead{Line}&\colhead{LTE}&\colhead{Virial}\\
   \colhead{Number} &\colhead{RA}&\colhead{Dec}&\colhead{Velocity}&&\colhead{Width}&\colhead{Mass}&\colhead{Mass}\\
&\colhead{($^{\circ}~^{\prime}~^{\prime\prime}$)}&\colhead{(d m s)}&\colhead{(km\,s$^{-1}$)}&\colhead{(pc)}&\colhead{(km\,s$^{-1}$)}&\colhead{(${\rm M}_\odot$)}&\colhead{(${\rm M}_\odot$)}
}
\startdata
01 & 03 29 02.6 & +31 14 56 & 7.80 & 0.02 & 0.62 & 2  & 2 \\
02 & 03 29 03.7 & +31 15 00 & 7.17 & 0.02 & 0.34 & 1  & 0.5 \\
03 & 03 29 03.2 & +31 15 18 & 7.64 & 0.02 & 0.54 & 2 & 1 \\
04 & 03 29 02.4 & +31 15 46 & 8.12 & 0.02 & 0.61 & 2 & 2 \\
05 & 03 29 02.6 & +31 15 54 & 8.27 & 0.03 & 0.70 & 2 & 3 \\
06 & 03 29 00.7 & +31 15 32 & 8.43 & 0.02 & 0.64 & 1 & 2 \\
07 & 03 29 06.8 & +31 15 44 & 8.12 & 0.03 & 0.70 & 3 & 3 \\
08 & 03 29 00.4 & +31 12 04 & 7.17 & 0.02 & 0.58 & 2 & 1 \\
09 & 03 29 00.1 & +31 12 18 & 7.17 & 0.02 & 0.56 & 1 & 1 \\
10 & 03 29 02.9 & +31 11 44 & 7.33 & 0.005 & 0.20 & 0.06 & 0.08 \\
11 & 03 28 55.9 & +31 14 16 & 7.33 & 0.03 & 0.56 & 2 & 2\\
12 & 03 28 57.0 & +31 13 56 & 7.02 & 0.03 & 0.39 & 1 & 1 \\
13 & 03 28 54.3 & +31 14 46 & 7.17 & 0.02 & 0.55 & 1 & 1 \\
14 & 03 29 04.9 & +31 20 58 & 7.64 & 0.02 & 0.52 & 1 & 1 \\
15 & 03 29 11.8 & +31 18 30 & 8.43 & 0.03 & 0.50 & 2 & 2 \\
16 & 03 29 05.4 & +31 16 12 & 8.43 & 0.01 & 0.31 & 0.4 & 0.2 \\
17 & 03 29 01.8 & +31 14 36 & 6.55 & 0.02 & 0.61 & 1 & 2 \\
18 & 03 29 04.2 & +31 14 48 & 6.70 & 0.02 & 0.37 & 0.9 & 0.6 \\
19 & 03 29 07.8 & +31 14 50 & 7.02 & 0.03 & 0.41 & 1 & 1 \\
20 & 03 28 36.2 & +31 13 24 & 7.33 & 0.01 & 0.34 & 0.7 & 0.2 \\
21 & 03 29 08.4 & +31 15 02 & 7.17 & 0.01 & 0.21 & 0.3 & 0.09 \\
22 & 03 29 09.0 & +31 15 10 & 7.33 & 0.02 & 0.59 & 2 & 2 \\
23 & 03 29 10.7 & +31 15 02 & 7.80 & 0.02 & 0.32 & 0.9 & 0.4 \\
24 & 03 29 08.8 & +31 14 44 & 7.49 & 0.02 & 0.37 & 1 & 0.6 \\
25 & 03 28 35.1 & +31 13 20 & 7.64 & 0.02 & 0.48 & 0.6 & 1 \\
26 & 03 29 02.3 & +31 20 30 & 7.80 & 0.02 & 0.61 & 2 & 2 \\
27 & 03 28 59.6 & +31 21 36 & 7.64 & 0.02 & 0.56 & 1 & 1 \\
28 & 03 29 13.1 & +31 13 54 & 7.80 & 0.02 & 0.44 & 1 & 0.8 \\
29 & 03 29 10.7 & +31 13 52 & 7.64 & 0.01 & 0.51 & 0.7 & 0.6 \\
30 & 03 28 40.3 & +31 17 40 & 7.80 & 0.03 & 0.42 & 2 & 1 \\
31 & 03 28 42.2 & +31 17 32 & 8.12 & 0.02 & 0.39 & 0.6 & 0.6 \\
32 & 03 28 55.3 & +31 19 16 & 7.80 & 0.02 & 0.45 & 0.8 & 0.9 \\
33 & 03 29 00.6 & +31 20 26 & 7.96 & 0.01 & 0.55 & 0.8 & 0.6 \\
34 & 03 28 59.2 & +31 20 22 & 7.64 & 0.01 & 0.31 & 0.3 & 0.2 \\
35 & 03 28 39.4 & +31 18 28 & 8.12 & 0.02 & 0.44 & 1 & 0.8 \\
36 & 03 28 53.9 & +31 14 54 & 8.27 & 0.01 & 0.41 & 0.6 & 0.4 \\
37 & 03 29 10.4 & +31 13 34 & 7.17 & 0.01 & 0.70 & 0.7 & 1 \\
38 & 03 29 11.8 & +31 13 08 & 6.86 & 0.02 & 0.58 & 0.5 & 1 \\
39 & 03 29 08.4 & +31 14 02 & 6.86 & 0.01 & 0.29 & 0.2 & 0.2 \\
40 & 03 29 07.8 & +31 14 10 & 6.86 & 0.01 & 0.33 & 0.3 & 0.2 \\
41 & 03 28 53.9 & +31 12 58 & 7.02 & 0.01 & 0.41 & 0.7 & 0.4 \\
42 & 03 29 14.9 & +31 12 44 & 7.33 & 0.01 & 0.44 & 0.6 & 0.4 \\
43 & 03 29 12.6 & +31 13 14 & 7.33 & 0.01 & 0.29 & 0.4 & 0.2 \\
44 & 03 29 09.0 & +31 14 10 & 7.17 & 0.01 & 0.22 & 0.2 & 0.1 \\
45 & 03 29 08.4 & +31 14 12 & 7.17 & 0.01 & 0.29 & 0.1 & 0.2 \\
46 & 03 29 16.5 & +31 12 06 & 7.49 & 0.01 & 0.26 & 0.1 & 0.1 \\
47 & 03 29 15.7 & +31 12 32 & 7.49 & 0.01 & 0.33 & 0.2 & 0.2 \\
48 & 03 29 13.8 & +31 13 14 & 7.49 & 0.02 & 0.34 & 0.4 & 0.5 \\
49 & 03 29 09.0 & +31 14 10 & 7.17 & 0.01 & 0.22 & 0.2 & 0.1 \\
50 & 03 28 49.7 & +31 14 30 & 7.49 & 0.02 & 0.44 & 0.4 & 0.8 \\
51 & 03 28 51.1 & +31 14 30 & 7.64 & 0.01 & 0.35 & 0.3 & 0.3 \\
52 & 03 29 00.7 & +31 15 26 & 7.49 & 0.01 & 0.21 & 0.3 & 0.09 \\
53 & 03 29 00.0 & +31 20 54 & 7.49 & 0.02 & 0.25 & 0.4 & 0.3 \\
54 & 03 28 51.8 & +31 15 34 & 7.64 & 0.01 & 0.27 & 0.3 & 0.2 \\
55 & 03 28 58.7 & +31 15 44 & 7.64 & 0.01 & 0.22 & 0.2 & 0.1 \\
56 & 03 28 53.6 & +31 18 24 & 7.64 & 0.03 & 0.53 & 0.9 & 2 \\
57 & 03 29 02.3 & +31 19 52 & 7.64 & 0.02 & 0.36 & 0.4 & 0.5 \\
58 & 03 29 04.0 & +31 19 22 & 8.12 & 0.01 & 0.45 & 0.3 & 0.4 \\
59 & 03 28 58.2 & +31 20 58 & 7.64 & 0.01 & 0.29 & 0.3 & 0.2 \\
60 & 03 28 47.3 & +31 15 18 & 7.96 & 0.02 & 0.30 & 0.4 & 0.4 \\
61 & 03 28 33.0 & +31 15 28 & 7.96 & 0.01 & 0.40 & 0.1 & 0.3 \\
62 & 03 28 49.2 & +31 16 06 & 8.12 & 0.01 & 0.33 & 0.3 & 0.2 \\
63 & 03 28 47.9 & +31 16 04 & 8.12 & 0.01 & 0.35 & 0.3 & 0.3 \\
64 & 03 29 08.7 & +31 17 32 & 8.27 & 0.02 & 0.37 & 0.5 & 0.6 \\
65 & 03 29 09.3 & +31 14 46 & 8.27 & 0.01 & 0.19 & 0.2 & 0.08 \\
66 & 03 29 09.2 & +31 16 54 & 8.27 & 0.02 & 0.38 & 0.7 & 0.6 \\
67 & 03 29 08.7 & +31 17 32 & 8.27 & 0.02 & 0.37 & 0.5 & 0.6 \\
68 & 03 29 07.1 & +31 17 28 & 8.43 & 0.02 & 0.43 & 0.5 & 0.8 \\
69 & 03 29 04.8 & +31 18 36 & 8.27 & 0.02 & 0.55 & 0.6 & 1 \\
70 & 03 28 57.1 & +31 22 04 & 8.59 & 0.01 & 0.40 & 0.2 & 0.3 \\
71 & 03 29 03.2 & +31 13 36 & 7.02 & 0.02 & 0.40 & 0.6 & 0.7 \\
72 & 03 28 50.3 & +31 11 34 & 7.02 & 0.003 & 0.36 & 0.05 & 0.3 \\
73 & 03 29 09.9 & +31 15 26 & 7.02 & 0.01 & 0.18 & 0.05 & 0.07 \\
74 & 03 29 04.8 & +31 11 56 & 7.17 & 0.005 & 0.17 & 0.05 & 0.06 \\
75 & 03 29 04.2 & +31 14 04 & 7.49 & 0.01 & 0.37 & 0.2 & 0.3 \\
76 & 03 29 09.0 & +31 21 06 & 7.33 & 0.01 & 0.28 & 0.08 & 0.2 \\
77 & 03 29 06.3 & +31 12 52 & 7.49 & 0.01 & 0.20 & 0.1 & 0.08 \\
78 & 03 28 47.5 & +31 14 24 & 7.64 & 0.007 & 0.23 & 0.2 & 0.1 \\
79 & 03 28 50.9 & +31 17 48 & 7.49 & 0.01 & 0.19 & 0.08 & 0.08 \\
80 & 03 29 18.4 & +31 11 42 & 7.64 & 0.01 & 0.16 & 0.08 & 0.05 \\
81 & 03 28 59.6 & +31 13 46 & 7.80 & 0.01 & 0.22 & 0.1 & 0.1 \\
82 & 03 29 13.4 & +31 14 36 & 7.64 & 0.01 & 0.30 & 0.1 & 0.2 \\
83 & 03 28 50.6 & +31 16 38 & 7.64 & 0.01 & 0.26 & 0.08 & 0.1 \\
84 & 03 29 06.0 & +31 16 42 & 8.12 & 0.02 & 0.34 & 0.4 & 0.5 \\
85 & 03 28 42.0 & +31 19 06 & 7.80 & 0.006 & 0.16 & 0.05 & 0.05 \\
86 & 03 29 06.7 & +31 19 50 & 7.96 & 0.006 & 0.17 & 0.05 & 0.06 \\
87 & 03 28 33.6 & +31 16 42 & 7.96 & 0.004 & 0.29 & 0.05 & 0.2 \\
88 & 03 28 34.2 & +31 16 46 & 7.96 & 0.01 & 0.33 & 0.2 & 0.2 \\
89 & 03 28 38.0 & +31 16 40 & 7.96 & 0.01 & 0.25 & 0.06 & 0.1 \\
90 & 03 28 33.6 & +31 15 40 & 8.27 & 0.01 & 0.16 & 0.08 & 0.05 \\
91 & 03 28 37.2 & +31 17 44 & 8.12 & 0.004 & 0.20 & 0.06 & 0.08 \\
92 & 03 28 57.0 & +31 19 42 & 8.27 & 0.01 & 0.32 & 0.1 & 0.2 \\
93 & 03 28 50.7 & +31 18 58 & 8.43 & 0.01 & 0.16 & 0.08 & 0.05 \\
\enddata
\label{n2hresults}

\end{deluxetable}

\clearpage

%
%
%
%

\end{document}